\documentclass[fleqn,usenatbib]{mnras}

\usepackage{newtxtext,newtxmath}

\usepackage[T1]{fontenc}
\usepackage{float}

\DeclareRobustCommand{\VAN}[3]{#2}
\let\VANthebibliography\thebibliography
\def\thebibliography{\DeclareRobustCommand{\VAN}[3]{##3}\VANthebibliography}

\usepackage{graphicx}	
\usepackage{amsmath}	

\usepackage{float}
\usepackage{subfig}

\def\lsim{\mathrel{\rlap{\lower4pt\hbox{\hskip1pt$\sim$}}
    \raise1pt\hbox{$<$}}}         
\def\gsim{\mathrel{\rlap{\lower4pt\hbox{\hskip1pt$\sim$}}
    \raise1pt\hbox{$>$}}}         

\title[Evolution of neutron star low mass X-ray binaries]{
Long-term evolution of spin and other properties of neutron star low-mass X-ray binaries: implications for millisecond X-ray pulsars}

\author[Kar and Ojha, et al.]{Abhijnan Kar $^{1}$$^{*}$\thanks{E-mail: karabhijnan123@gmail.com}, 
Pulkit Ojha$^{2}$$^{*}$\thanks{E-mail: pulkitojha0901@gmail.com} 
and Sudip Bhattacharyya$^{3}$\thanks{E-mail: sudip@tifr.res.in}
\\
$^{1}$Department of Physical Sciences, Indian Institute of Science Education and Research Berhampur, Vigyanpuri, Ganjam, 760003, India \\
$^{2}$School of Physical Sciences, National Institute of Science Education and Research Bhubaneswar, Jatani, Khurda, 752050, India \\
$^{3}$Department of Astronomy and Astrophysics, Tata Institute of Fundamental Research, 1 Homi Bhabha Road, Colaba, Mumbai 400005, India\\
$^{*}$\textbf{These two authors contributed equally to this paper and are designated as co-first authors}
}

\date{Accepted 2024 October 9. Received 2024 September 10; in original
form 2023 September 30}

\pubyear{2015}

\begin{document}
\label{firstpage}
\pagerange{\pageref{firstpage}--\pageref{lastpage}}
\maketitle

\begin{abstract}
A neutron star (NS) accreting matter from a companion star in a low-mass X-ray binary (LMXB) system can spin up to become a millisecond pulsar (MSP). Properties of many such MSP systems are known, which is excellent for probing fundamental aspects of NS physics when modelled using the theoretical computation of NS LMXB evolution. Here, we systematically compute the long-term evolution of NS, binary and companion parameters for NS LMXBs using the stellar evolution code MESA. 
We consider the baryonic to gravitational mass conversion to calculate the NS mass evolution and show its cruciality for the realistic computation of some parameters. With computations using many combinations of parameter values, we find the general nature of the complex NS spin frequency ($\nu$) evolution, which depends on various parameters, including accretion rate, fractional mass loss from the system, and companion star magnetic braking. 
Further, we utilize our results to precisely match some main observed  parameters,  such as $\nu$,  orbital period ($P_{\rm orb}$), etc., of four accreting millisecond X-ray pulsars (AMXPs). 
By providing the $\nu$, $P_{\rm orb}$ and the companion mass spaces for NS LMXB evolution, we indicate the distribution and plausible evolution of a few other AMXPs. 
We also discuss the current challenges in explaining the parameters of AMXP sources with brown dwarf companions and indicate the importance of modelling the transient accretion in LMXBs as a possible solution.
\end{abstract}




\section{Introduction}


A neutron star (NS) low-mass X-ray binary (LMXB) is a binary stellar system, in which the NS accretes matter from a 
low-mass ($\lsim 1.5 M_\odot$) companion or donor star, when the latter fills its Roche lobe \citep[RL; ][]{review_bhattacharya}. The magnetic field of an NS in such a system is decayed to a lower value of $\sim 10^{8-9}$ G, and the lifetime of an LMXB is typically $\gsim 1$ Gyr.
Such a system primarily emits X-rays from the inner part of its accretion disc, through which matter is transferred to the NS, as well as from the NS surface. The high specific angular momentum of the accreting matter from the disc can spin up the NS to frequencies of 
several hundred Hz. 
A few of such rapidly-spinning NSs can be observed as millisecond pulsars (MSPs). 
These recycled MSPs with 
low magnetic field and high spin frequency occupy a distinctly different region in the NS magnetic field ($B$) versus NS spin period ($P$) plot relative to the more numerous slowly spinning pulsars 
\citep{review_bhattacharya}.
Many of the NS LMXBs, having magnetic field strong  to channel the accreted matter to the magnetic poles, show periodic X-ray intensity variation at the NS spin frequency ($\nu$), and hence are called accreting millisecond X-ray pulsars \citep[AMXPs; ][]{patruno_amxp,Salvo2022}.When accretion stops at the end of the LMXB phase, we observe many such NSs as radio MSPs \citep{Bhattacharyya_2021}.



In this paper, we aim to study the evolution trends of several NS LMXB parameters, including $\nu$.
More than 20 AMXPs \citep{Salvo2022} and hundreds of radio MSPs have been observed. 
Their properties, such as
binary orbital period, companion star mass, NS spin frequency and magnetic field, etc., are publicly
available\footnote{E.g., \url{http://www.atnf.csiro.au/research/pulsar/psrcat/}.} in many cases. However, their evolution depends on various aspects of binary evolution of their progenitors, such as the accretion through the NS magnetosphere in a strong gravity region, the NS structure, and gravitational radiation. Thus, the computation of NS evolution through the LMXB phase can be useful to probe these fundamental aspects of NS physics.
We compute LMXB evolution using the “Modules for Experiments in Stellar Astrophysics” (MESA) code \citep{mesa2011}. We particularly probe how the NS spin frequency evolves through the LMXB phase, useful to understand the known spin frequency distribution of radio and X-ray MSPs \citep{Bhattacharyya_2021, Salvo2022} .
Before MESA, several authors explored the binary evolution of LMXBs and its impact on the formation of MSPs. There are three sets of parameters for NS LMXB systems, the orbital parameters, the companion star parameters and the NS parameters. 
\citet{podsiadlowski} indicated the binary parameter space of LMXBs and IMXBs for which the system would show stable mass transfer and different mass mechanisms would prevail. \citet{weimin} was one of the early papers to carefully explore the binary parameter space for which LMXBs can evolve into radio MSPs with He WD companions. Shortly afterwards, \citet{Tauris_2012_rldp} described all the possible cases of mass transfer scenarios in LMXBs and IMXBs, the effect of different parameters such as magnetic braking index and RL decoupling (RLDP) to obtain the parameters of a few observed radio MSPs in the $\nu$ vs $P_{\rm orb}$ plot. In another paper, \citet{istrate}, they also explored the effect of the evolution of LMXBs below the bifurcation period to obtain parameters of radio MSPs in tight binaries.
Development of MESA made the grid calculations of binaries convenient and authors such as \citet{Rapapport_grid_mesa} utilised it to explain radio MSP sources by considering a huge grid of progenitor LMXBs. Formation of redback and black widow MSPs was explored using MESA by \citet{Chen_2013}. Further, the formation of redbacks MSPs was also explained by \citet{main_jia&li,jia&li_beta}  including the effect of evaporation in the post-LMXB phase. A few other aspects of LMXB/IMXB evolution, such as the formation of ultracompact binaries were also studied using MESA  \citep[e.g., see ][]{Chen_2016, 10.1093/mnras/stab670}.
  

Although authors focused on various aspects of LMXB evolution, mainly orbital and companion star parameters, the evolution of NS parameters with binary ones, particularly using MESA, is poorly explored so far. A few papers, such as \citet{Tauris_2012_rldp, Tauris-science} considered the spin evolution of NS LMXBs in their grid calculations, though their results were focused towards explaining the radio MSPs formed in the post-LMXB phase. Further, a few recent papers, such as \citet{accreted_mass,magnetic_inclination} used MESA to compute binary evolutionary tracks and explore a few properties of the NS, including spin frequency. However, \citet{magnetic_inclination} focused mainly on magnetic inclination evolution with various initial parameters, while \citet{accreted_mass} explored maximum accreted mass onto NS and its effect on spin frequency evolution. \citet{sudip_2017_2, sudip2021} considered the spin evolution of NS, modelling the accretion phases of transient LMXBs, but the evolution of binary parameters was not calculated. In this paper, we consider a systematic variation of all the parameters of NS LMXBs, their effects on the evolution of NS spin frequency, and all the other relevant NS, companion star, and binary parameters. 

%
In contrast to many works reproducing the parameters of observed radio MSP sources, only a few papers, such as \cite{xianghe} and \cite{amxp_suvorov}, attempted to reproduce the parameters of AMXPs through long-term binary evolution computations. \citet{xianghe} matched a few observed parameters for AMXPs such as companion mass and orbital period, 
but not $\nu$.
\citet{amxp_suvorov} considered different torque mechanisms to explain spin-up rates of a few observed AMXPs, but did not consider the long-term accretion of the binary. 
In this paper, we use our general results of LMXB parameter evolution to show that the observed parameters, such as companion mass, orbital period, NS spin frequency and limiting value of NS magnetic field, can be precisely obtained for a few AMXP sources. 
Overall, we provide the distribution of AMXP sources in the relevant parameter space including the spin evolution of the NS, which could not be found in the existing literature to the best of our knowledge.



This paper is structured as follows. We discuss the theoretical background of various aspects of LMXB evolution, along with any new method, in section~\ref{sec: methods}. Various general results of LMXB evolution and the corresponding figures are presented 
in section~\ref{sec: Results}. In section \ref{sec:AMXPs}, we discuss the evolution scenario of various AMXPs and how can their observed parameters, including spin frequency, be obtained from our general results. We discuss the implications of the obtained results in section~\ref{sec:discussion}. Finally, a summary of the important findings and conclusions is provided in section~\ref{sec:summary}.

\section{Methods}
\label{sec: methods}
\subsection{Modelling LMXB evolution}
\label{subsec: medlling LMXB}
\subsubsection{Binary evolution setup}
 
``Modules for Experiments in Stellar Astrophysics'' \cite[MESA; ][]{mesa2011,mesa2013,mesa2015,mesa2018,mesa2019} is an open-source 1D stellar evolution code\footnote{\url{https://docs.mesastar.org/en/release-r23.05.1/}}. 
Its primary purpose is to facilitate the comprehensive exploration of stars and their lifecycle, spanning from the initial stages to the more advanced phases of evolution, as well as evolution of binary star systems and its various properties. 
Here, we use the MESA version r22.05.01 , specifically its binary module to compute the evolution of LMXB systems.
In our computation, the LMXB initially consists of a NS and a zero age main-sequence (ZAMS) companion/companion star with the solar composition \citep[e.g., ][]{main_jia&li}.
The NS is considered to be a point mass ($M_{\rm NS}$) of 1.35 $M_{\odot}$ initially \citep[e.g., ][]{main_paper_sudip} and the companion/donor mass ($M_{\rm comp}$) is taken in the range of 0.5 $M_{\odot}$ to 1.5 $M_{\odot}$. 
The Roche Lobe radius of the companion star ($R_{\rm L,comp}$) is given by \citep{eggleton}:
\begin{equation}\label{eq 1}
\frac{R_{\rm L,comp}}{a} = \frac{0.49 q^{-\frac{2}{3}}}{0.6 q^{-\frac{2}{3}} + \ln(1 + q^{-\frac{1}{3}})}, 
\end{equation}
where $q = \frac{M_{\rm NS}}{M_{\rm comp}}$, and $a$ is the orbital separation. The mode of accretion prevalent in LMXB systems is through Roche Lobe Overflow (RLOF), and the corresponding mass transfer rate is calculated using the expression from \citep{Ritter}: 
\begin{equation}\label{eq 2}
-\dot M_{\rm comp} = \dot{M_0}\exp\Bigl[\frac{R_{\rm comp} - R_{\rm L,comp}}{H_{\rm P}/\gamma(q)}\Bigr],
\end{equation}
where  $\dot{M}_{\rm comp}$ is the  mass loss rate from the companion star and hence is a negative quantity. Also, $\dot{M_0}$ depends on various parameters, $R_{\rm comp}$ is the radius and $H_{\rm P}$ is the pressure scale height of the atmosphere of the companion star, respectively, and $\gamma$ is a function of $q$.
For our systematic study of NS LMXB evolution, we use a default or canonical value and a range of values of each input parameter (see Table~\ref{table1}).
Note that angular momentum ($J$) loss (AML) mechanisms play an important role to drive the LMXB evolution. 

\subsubsection{Angular momentum Loss}\label{section_AML}

In this paper, we consider three most commonly used AML mechanisms: 
the gravitational radiation (GR) from the system;
magnetic braking (MB) of the companion; and mass loss (ML) from the system. 
Therefore, 
\begin{equation}\label{eq 3}
    \dot J = \dot{J}_{\rm GR} + \dot{J}_{\rm MB} + \dot{J}_{\rm ML}.
\end{equation}
The orbital angular momentum loss rate due to the gravitational radiation is \citep{Landau}
\begin{equation}\label{eq 4}
\dot{J}_{\rm GR} = -\frac{32}{5}\frac{G^{3.5}}{c^5} \frac{M^2_{\rm NS} M^2_{\rm comp} (M_{\rm NS} + M_{\rm comp})^{0.5}}{a^{3.5}},
\end{equation}
where $a$ is the separation between the binary components,  $G$ is the Gravitational constant and $c$ is the speed of light in vacuum.
The AML due to magnetic braking of the companion star is given by \citep{rapapport_mb}

\begin{equation}\label{eq 5}
\dot{J}_{\rm MB} = -3.8 \times 10^{-30} M_{\rm comp} R_{\rm comp}^{\gamma} \Omega^3 {\rm dyn~cm}.
\end{equation}
Here, $\gamma$ is the magnetic braking index, set to 3 by default in MESA. 
We also set $\gamma =  4$ \citep[see ][]{rapapport_mb,chen}
and check the difference in the evolutionary tracks. 
The angular spin frequency of the companion star, $\Omega$, should be equal to the orbital angular frequency due to the tidal synchronization. 

Magnetic braking becomes too small if the convective envelope becomes too thin in the course of the stellar evolution. Therefore, following  \citet{podsiadlowski,main_jia&li}, an ad hoc factor of 
$$ \exp\Bigl(\frac{-0.02}{q_{\rm conv}} + 1\Bigr) $$
is included.
Here,  $ q_{\rm conv} ($< 0.02$)$ is the convective mass fraction of the companion star.
We consider two cases regarding magnetic braking: (i) we switch it off when the star becomes fully convective  \citep{rapapport_mb,main_jia&li} and (ii) we keep it always on \citep{tailo,goodwin}.

The AML due to the mass loss from the system is given by \citep[e.g., ][]{mesa2015,goodwin}
\begin{equation}\label{eq 6}
\dot{J}_{\rm ML} = \beta \dot{M}_{\rm comp} \Bigl(\frac{M_{\rm comp}}{M_{\rm NS} + M_{\rm comp}}\Bigr)^2a^2 \Omega,
\end{equation}
where $\beta$ is the fractional loss (from the binary system) of the mass transferred from the companion star.
We use a range of values for $\beta$ with $\beta = 0.5$ as the canonical value (see Table \ref{table1}).


\subsection{Mass accretion onto NS}
\label{subsec: Mass Accretion}

The accretion rate ($\dot M$) onto the NS is limited by the Eddington limit given as \citet{main_jia&li}
\begin{equation}\label{eq 7}
 \dot{M}_{\rm Edd} = 3.6 \times 10^{-8} \Bigl(\frac{M_{\rm NS}}{1.4 M_{\odot}}\Bigr)\Bigl(\frac{0.1}{GM_{\rm NS}/R_{\rm NS}c^2}\Bigr)\Bigl(\frac{1.7}{1 + X}\Bigr)~{\rm M}_{\odot}~{\rm yr}^{-1},
\end{equation}
where $X$ is the hydrogen fraction and $R_{\rm NS}$ is the radius of the NS. 
Apart from this limit, $\dot M$ is also limited due to $\beta$.

We also calculate a critical accretion rate ($\dot{M}_{\rm crit}$) following \citet{King_mcrit,Lasota-mcrit}, below which the accretion disc should become thermally and viscously unstable, making the accretion onto the NS transient or episodic.
The $\dot{M}_{\rm crit}$ can be written as (\citet{sudip2021}) : 

\begin{equation}\label{eq 8}
\dot{M}_{\rm crit} \approx 3.2 \times 10^{15} \Bigl(\frac{M_{\rm NS}}{M_{\odot}}\Bigr)^{ \frac{2}{3}}\Bigl(\frac{P_{\rm orb}}{3~{\rm hr}}\Bigr)^{ \frac{4}{3}} {\rm g}~{\rm s}^{-1},
\end{equation}
where $P_{\rm orb}$ is the binary orbital period in hours. However, we do not consider transient accretion in this paper.

In our computation, we include irradiation of the companion star due to X-ray luminosity, which is given by
\begin{equation}\label{eq 9}
    L_{X} = \frac{GM_{\rm NS}\dot M}{R_{\rm NS}}.
\end{equation}
The corresponding irradiation flux is given by:
\begin{equation}\label{eq 10}
    F_{\rm irr} = \epsilon\frac{L_X}{4\pi a^2}.
\end{equation}
In our default model, we take column depth to be 10 g cm$^{-2}$, irradiation efficiency $\epsilon$ to be 0.015, $a$ to be $1\times10^{11}$ cm and limit the maximum irradiation flux to $3\times 10^9$ erg s$^{-1}$cm$^{-2}$, following \citet{goodwin}. We also explore some cases with higher $\epsilon$ values and the case without irradiation as well.

\subsection{Baryonic to gravitational mass conversion for NS}\label{Baryonic_gravitational}

The accretion rate ($\dot M$) is the baryonic/inertial mass of the matter transferred per unit time.
But the neutron star mass $M_{\rm NS}$ affecting the binary parameters is the gravitational mass.
When the accreted matter falls onto the NS, one needs to convert the baryonic mass into the gravitational mass to estimate the increased $M_{\rm NS}$. Typically, this is not done throughout the binary evolution, and hence there can be a substantial error in the estimation of $M_{\rm NS}$ evolution because NS is a very compact object \citep[e.g., ][]{manjari_bagchi}. 
A few previous papers, such as \citet{istrate, chen_conversion, Van_2019}, included a constant conversion factor between the gravitational mass and the baryonic mass. 
However, in this paper, we change the conversion factor dynamically, and hence realistically, during the accretion process,
because here we include this effect by doing the conversion at every step of the evolution,
using  Eq 19 of \citet{cipoletta}:
\begin{equation}\label{eq 11}
\begin{aligned}
M_G \approx M_{\rm B} - (1/20)M_{\rm B}^2 \\
M_{\rm B} \approx M_{\rm G} + (13/200)M_{\rm G}^2.
\end{aligned}
\end{equation}
Here, $M_{\rm G}$ ($= M_{\rm NS}$) is the gravitational mass and $M_{\rm B}$ is the baryonic mass of the NS. 
To modify the  accretion rate according to the above prescription, we use the time derivative of the first equation to express $\dot{M_{\rm G}}$
in terms of $\dot{M_{\rm B}}$ (calculated by MESA) as 
\begin{equation}\label{eq 12}
\dot{M}_{\rm G} = (1 - 0.1(M_{\rm G} +  \frac{13}{200}M_{\rm G}^2))\dot{M}_{\rm B}
\end{equation}
where $\dot M_{\rm G}$ is the rate of change of gravitational mass and $\dot M_{\rm B}$ is the rate at which baryonic mass is accreted. The accreted gravitational mass  (which is simply the rate of change in neutron star gravitational mass $M_{\rm G}$) gets added to the NS mass in MESA. 
This conversion factor effectively acts as another efficiency constraint on the rate of accretion, as discussed later.

\subsection{NS Spin evolution}
\label{subsec:spin evolution} 

To calculate the accretion-driven spin evolution of the NS in an LMXB system, we follow the torque prescription described in \citep{main_paper_sudip}. 
The gravitational mass, radius, spin frequency, and magnetic dipole moment of NS are  
$M_{\rm NS}$, $R_{\rm NS}$, $\nu$, and $\mu = B R_{\rm NS}^3$, respectively, where $B$ is the surface dipole magnetic field 
\citep{main_paper_sudip}. Following \citet{shibazakiB}, we evolve the NS magnetic field as 
\begin{equation}\label{B eq}
    B_{\rm f} = \frac{B_{\rm i}}{1 + \Delta M_{\rm acc}/m_{\rm B}}
\end{equation}
where $m_{\rm B}$ $\sim$~ 10$^{-4} M_{\rm \odot}$.
We take $B_{\rm i}$ to be in order of $\sim 10^{11} - 10^{12}$ G, such that the decayed magnetic field, $B_{\rm f}$, remains in order of $10^8$ G.
We also calculate NS radius as $R_{\rm NS} = AM_{\rm NS}^{1/3}$, where the value of the proportionality constant $A$ has been fixed by taking an NS radius of 11.2 km for the NS mass of $1.4 \rm M_{\odot}$.
There are three 
characteristic radii related to the accretion process. 
The accretion disc is extended up to the 
magnetospheric radius ($r_{\rm m}$), inside which the NS magnetosphere channels accreted matter onto
the stellar magnetic polar caps. 
This radius is given by
\begin{equation}\label{eq 13}
r_{\rm m} = \xi\Bigl(\frac{\mu^4}{2G M_{\rm NS}{\dot{M}}^2}\Bigr)^{\frac{1}{7}}, 
\end{equation} 
where $\xi$ is a constant of order unity, and we consider a value of 1 for the purpose of demonstration.
The corotation radius ($r_{\rm co}$), the distance at which the Keplerian angular velocity of the material in the accretion disc is equal to the NS spin angular velocity, is given by 
\begin{equation}\label{eq 14}
r_{\rm co} = \Bigl(\frac{G M_{\rm NS}}{4\pi^2\nu^2}\Bigr)^{\frac{1}{3}}.
\end{equation}
Finally, the light-cylinder radius ($r_{\rm lc}$), the distance at which the linear velocity of rotating magnetic field lines is equal to the speed of light in vacuum, is given by 
\begin{equation}\label{eq 15}
  r_{\rm lc} = \frac{c}{2 \pi \nu}.  
\end{equation}
The relative values of these characteristic radii indicate two phases of accretion -- the accretion phase ($r_{\rm m} < r_{\rm co}$), and the propeller phase ($r_{\rm co} < r_{\rm m} < r_{\rm lc}$). 
Steady accretion happens when the accretion disc extends till inside $r_{\rm co}$ \citep[i.e., $r_{\rm m} < r_{\rm co}$; e.g., ][]{ghosh_rmag}. 
In the case of the `propeller phase', accretion is largely shut off by a centrifugal barrier  \citep{sunayev}.
Note that $r_{\rm m}$ cannot be smaller than the radius ($r_{\rm ISCO}$) of the innermost stable circular orbit (ISCO) or $R_{\rm NS}$, whichever is larger.


To model the NS spin frequency evolution, we consider two components of the accretion-related torque -- one due to the angular momentum change of the accreted matter and another due to the interaction of NS magnetic field lines with the disc.
As calculated by \citet{Rappaport_2004, 
main_paper_sudip}, the expression of the torque in the accretion phase is given as 
\begin{equation}\label{eq 19}
N = \dot{M}\sqrt{GM_{\rm NS}r_{\rm m}} + \frac{\mu^2}{9r_{\rm m}^3}\Biggl[2\Bigl(\frac{r_{\rm m}}{r_{\rm co}}\Bigr)^3 - 6\Bigl(\frac{r_{\rm m}}{r_{\rm co}}\Bigr)^{\frac{3}{2}} + 3\Biggr],
\end{equation}
and that in the propeller phase is given by
\begin{equation}\label{eq 20}
N = - \eta\dot{M}\sqrt{GM_{\rm NS}r_{\rm m}} - \frac{\mu^2}{9r_{\rm m}^3}\Biggl[3 - 2\Bigl(\frac{r_{\rm co}}{r_{\rm m}}\Bigr)^{3/2}\Biggr],
\end{equation}
where $\eta$ is an order of unity constant. 
These expressions of the torque show that the NS should spin up in the accretion phase and spin down in the propeller phase. 
Since for a constant $\dot M$, $r_{\rm co}$ decreases by spin-up when $r_{\rm co} > r_{\rm m}$, and $r_{\rm co}$ increases by spin-down when $r_{\rm co} < r_{\rm m}$, the NS should approach a spin equilibrium condition of $r_{\rm co} = r_{\rm m}$, and the corresponding equilibrium spin frequency is 
\begin{equation}\label{eq 23}
\nu_{\rm eq} = \frac{1}{2\pi}\sqrt{\frac{GM_{\rm NS}}{r_{\rm m}^3}} = \frac{1}{2^{11/14}\pi\xi^{3/2}}\Bigl(\frac{G^5M_{\rm NS}^5\dot{M}^3}{\mu^6}\Bigr)^{1/7}.
\end{equation}


Note that, for transient/episodic accretion  (for $\dot{M} < \dot{M}_{\rm crit}$; section~\ref{subsec: Mass Accretion}) , the equilibrium spin frequency is expected to be a few hundred Hz higher, and $\nu$ should approach this higher equilibrium frequency  \citep{main_paper_sudip,sudip_2017_2,sudip2021}  
However, in this paper, we study NS spin evolution only for non-episodic accretion.

For each step of MESA, change in the angular momentum of the NS is calculated from 
\begin{equation}\label{eq 21}
  \Delta J = N \Delta t,
\end{equation}
where $\Delta t$ is the time step.
Total angular momentum of the NS is updated at every step by adding $\Delta J$ to the previous value of $J$. Then, the new NS spin frequency is  calculated from 
\begin{equation}\label{eq 22}
  \nu = \frac{J}{2\pi I},
\end{equation}
where I is the moment of inertia of the NS and can be written as $AM_{\rm NS}R_{\rm NS}^2$. 
The coefficient $A$ should typically be in the range of $0.33-0.43$ \citep{sudip_newastron}
but we use a range of $0.33-0.5$ here (Table~\ref{table1}) to test NS evolution for higher $I$ values.

Apart from the accretion torques, we also compute evolution for an additional spin-down torque ($N_{\rm GW}$) due to the emission of gravitational waves from the NS \citep{Bildsten_1998}:  
\begin{equation}\label{eq 24}
N_{\rm GW} = -\frac{32GQ^2}{5}\Bigl(\frac{2\pi\nu}{c}\Bigr)^5,
\end{equation}
where Q is the misaligned mass quadrupole moment of the NS. We consider a range of values for Q in this paper \citep{ellipcity_quadrupole}.
In Table \ref{table1}, we list all the input parameters for MESA computation, and their default/canonical values and ranges of values.

\begin{table*}
\begin{center}
\caption{A list of input parameters with default/canonical values and ranges of values used for computation of NS LMXB evolution using the MESA code (see section~\ref{sec: methods}).}
\begin{tabular}{lcc} 
    \hline
Parameter & Canonical value & Used range  \\
\hline
Initial orbital period $P_{\rm orb}$ (day) & 1.0 & $0.5-2.5$ \\
Initial companion mass $M_{\rm comp}$ ($M_{\odot}$) & $1.0$ & $0.5-1.5$ \\
Initial NS spin frequency $\nu$ (hz) & 1 & $1-100$ \\
\hline
Fractional mass loss $\beta$ & 0.5 & $0.1-0.9$ \\
Magnetic braking Index & 3 & 3 and 4 \\
$A$ of NS moment of inertia $AM_{\rm NS}R_{\rm NS}^2$ & 0.4 & $0.33 - 0.5$ \\
NS Mass quadrupole moment ($10^{36}$ g~cm$^2$) & 0 & $0.0 - 190.8$ \\
$\eta$ & 1 & $0.2 - 1$ \\
\hline
\end{tabular}
\label{table1}
\end{center}
\end{table*}

\section{Results}
\label{sec: Results}
\subsection{General aspects}
\label{subsec: general example}
\begin{figure*}
\centering
\includegraphics[width = 95 mm, scale = 2.0]{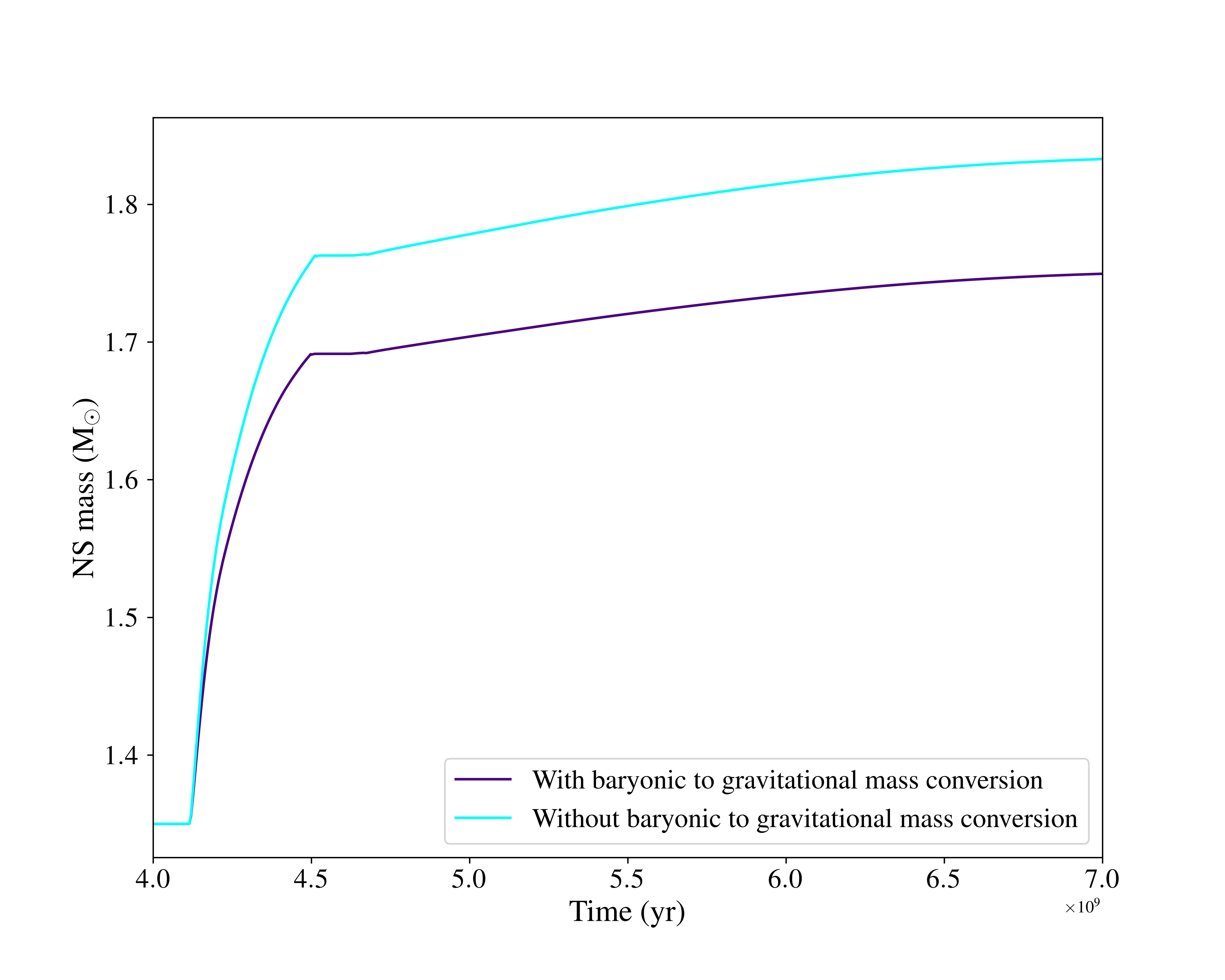}
\includegraphics[width = 95 mm, scale = 2.0]{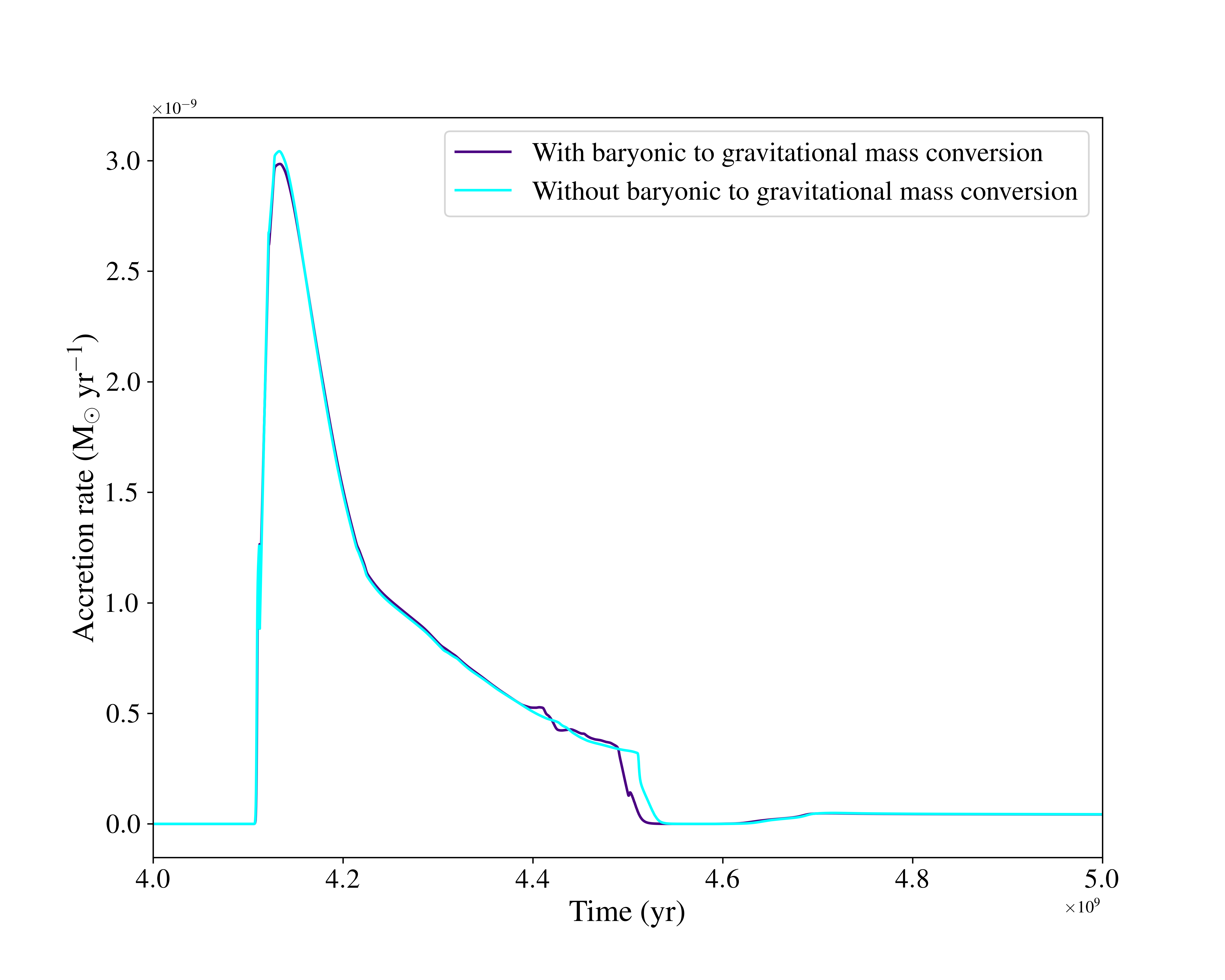}
\includegraphics[width = 95 mm, scale = 2.0]{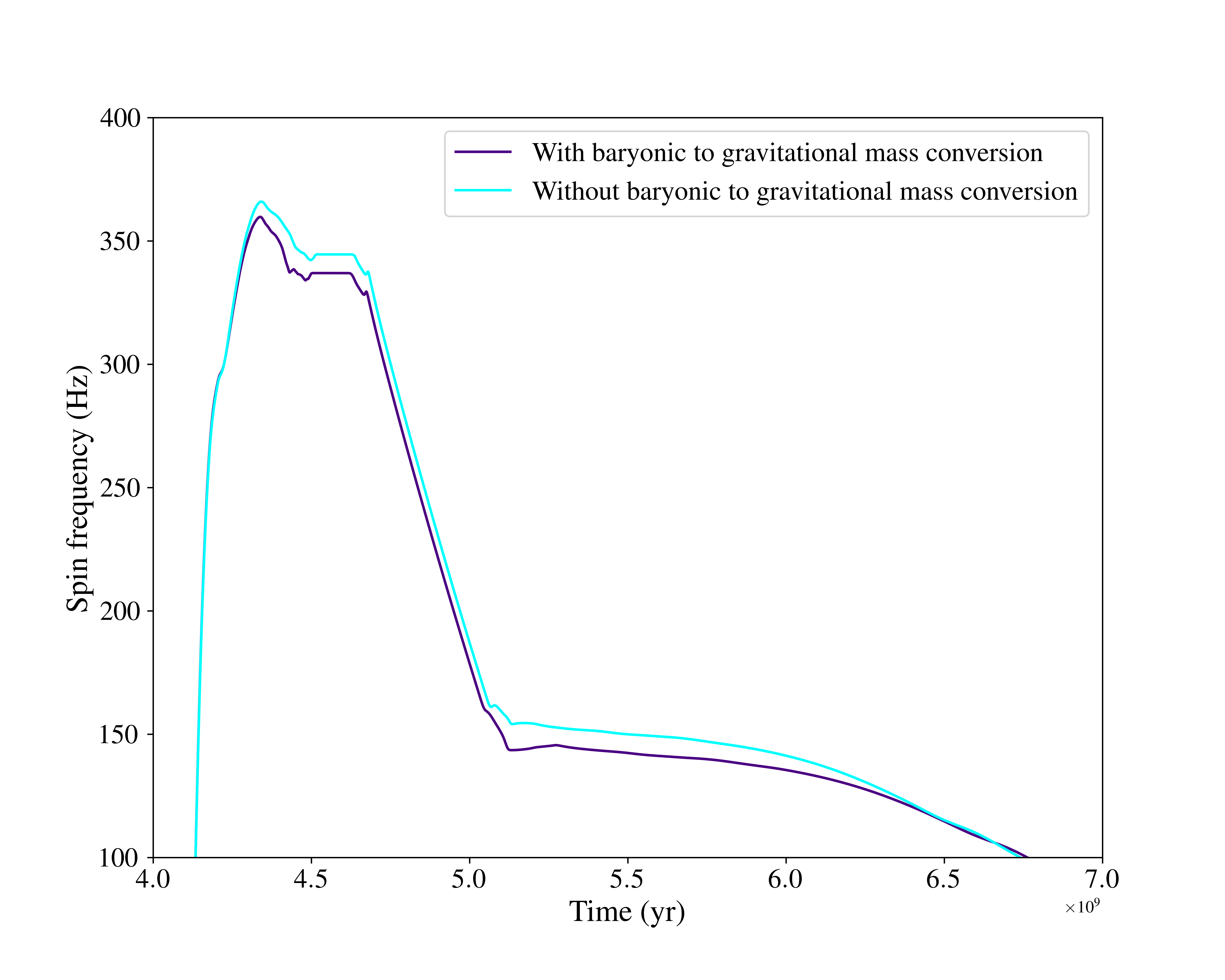}
\caption{Evolution of three parameters (NS mass, accretion rate and NS spin frequency) of a NS LMXB, 
as calculated using MESA. 
In each panel,
the yellow curve considers baryonic to gravitational mass conversion for computing NS mass evolution, while the blue curve does not consider this conversion.
Here, we assume an initial orbital period 
 of
$P_{\rm orb} = 1.7$ day and initial companion mass of 1.0 $\rm M_{\rm \odot}$. Other parameters are fixed at their canonical values (see Table~\ref{table1}).  
These panels show the effects of the conversion of baryonic to gravitational mass in the computation of NS mass (see section~\ref{subsec: general example}).
\\
}
\label{cipoletta}
\end{figure*}

Using MESA, we vary one parameter at a time and study the trends in the evolution of all kinds of NS LMXB parameters, i.e., binary, companion and NS parameters
With the help of the general results of  evolution of these parameters throughout the LMXB phase, we focus on obtaining the parameter space for accreting AMXPs.
To the best of our knowledge, we employ a way of the dynamic conversion of baryonic mass to gravitational for accretion rate calculation (see section~\ref{Baryonic_gravitational}). 
As Figure~\ref{cipoletta} shows, this effect is substantial for $M_{\rm NS}$, thus significantly affecting the evolution of $\dot M$ and $\nu$. We show that this effect should be included for a realistic computation of the long-term evolution of NS $\nu$ and other properties.

\begin{figure*}
\centering
\includegraphics[width= 90 mm, scale = 1]{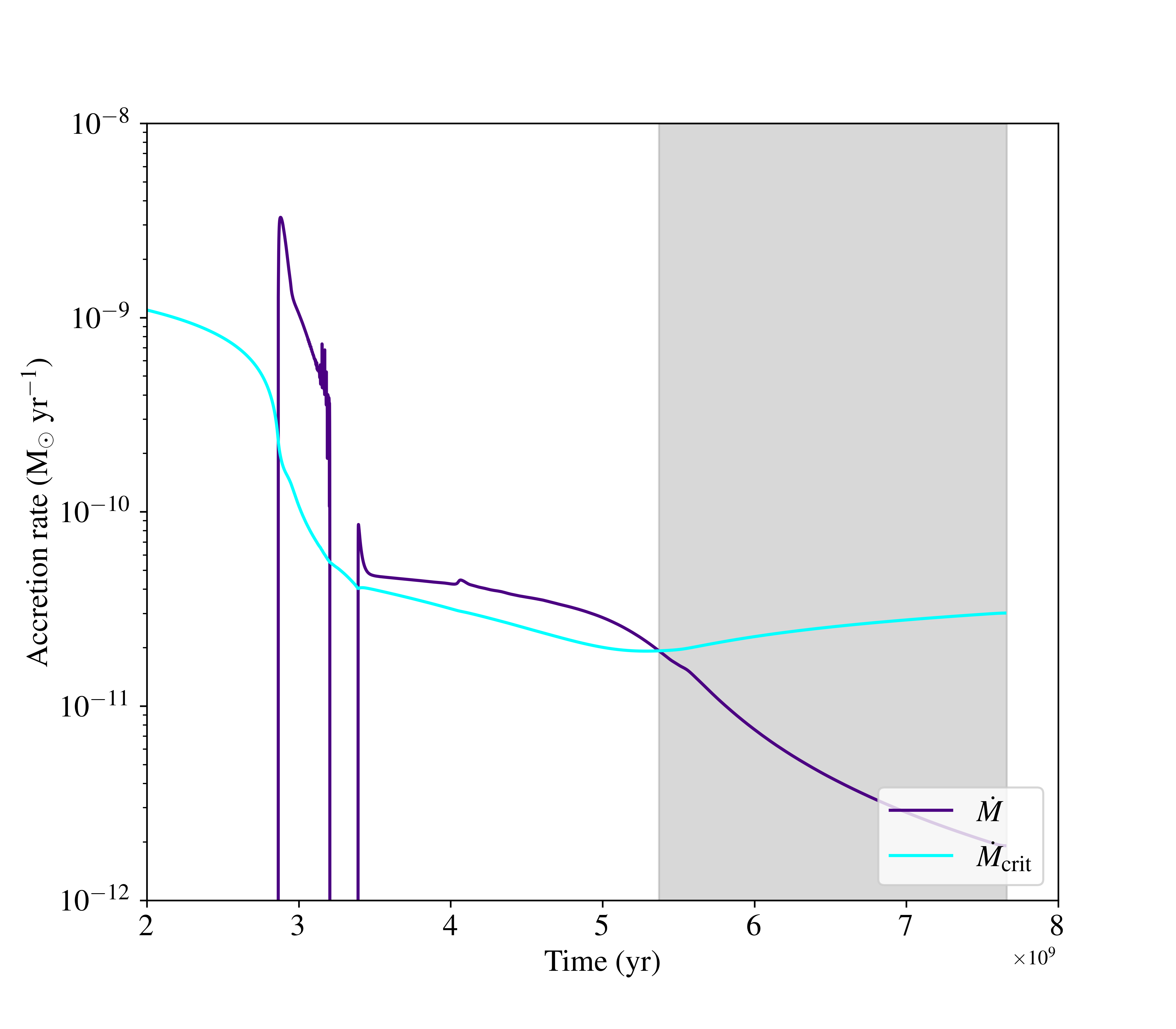}e
\includegraphics[width= 90 mm, scale = 1]{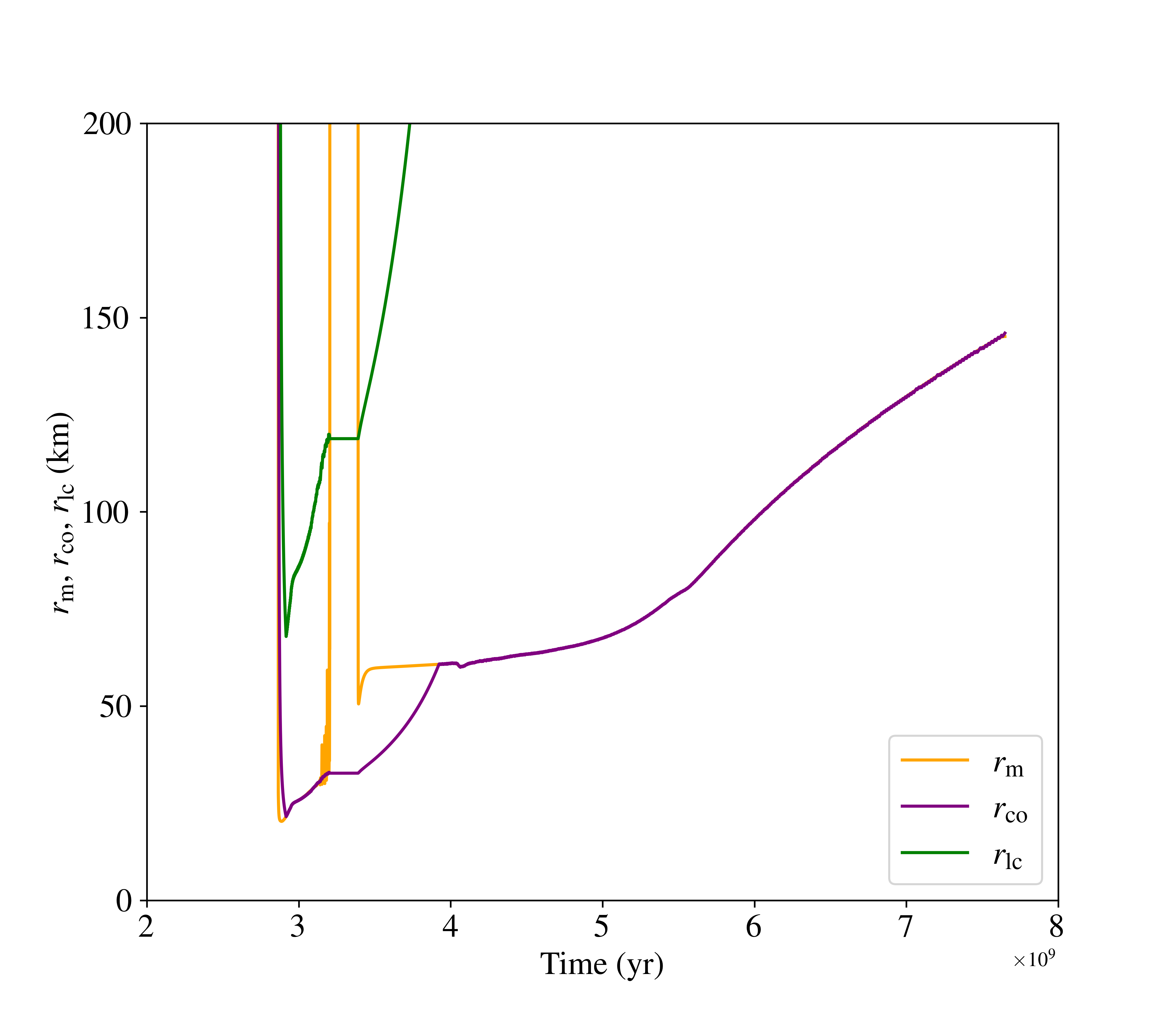}
\includegraphics[width= 90 mm, scale = 1]{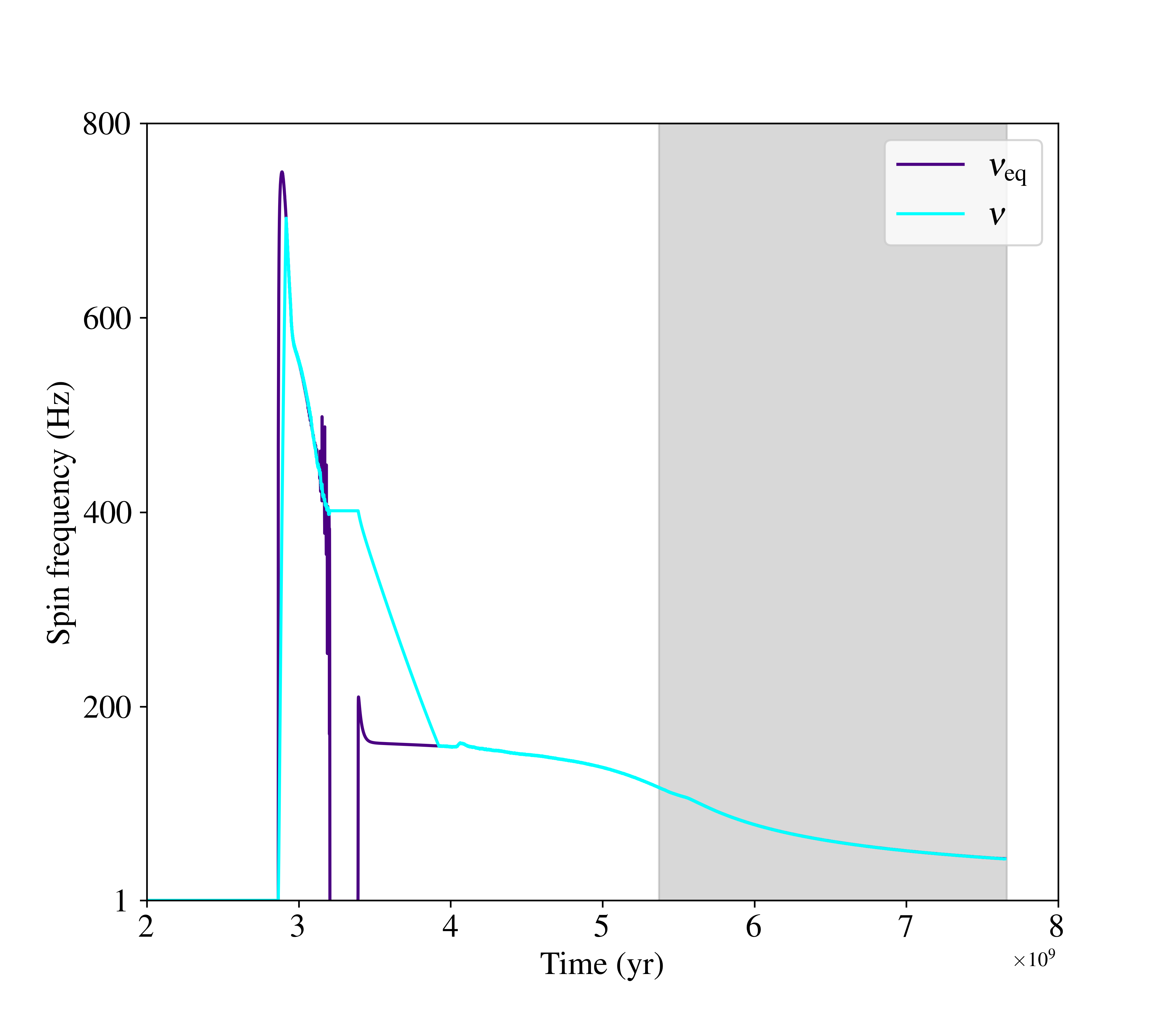}
\caption{Evolution of parameters (accretion rate [$\dot M$], critical accretion rate [$\dot M_{\rm crit}$], characteristic radii [$r_{\rm m}$, $r_{\rm co}$, $r_{\rm lc}$], spin frequencies [$\nu$ and $\nu_{\rm eq}$]) of an NS LMXB, as calculated using MESA.
Here, we assume an initial orbital period 
 of
$1.7$ day and initial companion mass of 1.0 $M_{\rm \odot}$. Other parameters are fixed at their canonical values (see Table~\ref{table1}).
The grey patch marked in panels of accretion rate and spin frequency denotes the transient phase of an LMXB evolution ($\dot{M} < \dot{M}_{\rm crit}$ ).
These panels show that the $\dot{M}$ evolution can be complex, which leads to a complex $\nu$ evolution (see section~\ref{subsec: general example}).}
\label{Consistency Check}
\end{figure*}

Figure~\ref{Consistency Check}, in three panels, shows an example of the evolution of $\dot M$, $\dot M_{\rm crit}$, $r_{\rm m}$, $r_{\rm co}$, $r_{\rm lc}$, $\nu$ and $\nu_{\rm eq}$ of an NS LMXB. 
The accretion rate panel shows that initially there is no accretion.
As the companion star fills its Roche lobe, $\dot M$ quickly attains a high value, and then decreases as the source parameters (e.g., masses and separation of two stellar components, companion star radius) evolve (Equations~\ref{eq 1} and \ref{eq 2}).
As $\dot M$ increases, $r_{\rm m}$ decreases (Equation~\ref{eq 13}) and the NS spins up, increasing $\nu$ and decreasing $r_{\rm co}$.
As $r_{\rm co}$ tends to $r_{\rm m}$, $\nu$ approaches  $\nu_{\rm eq}$ (Equation~\ref{eq 23}) and the NS attains the spin equilibrium (see section~\ref{subsec:spin evolution}).
However, as $\dot M$ subsequently decreases in course of the binary evolution, $\nu_{\rm eq}$ decreases, and $\nu$, which tracks $\nu_{\rm eq}$, also decreases. Thus, for this particular example of initial parameters, $\nu$ has a high value ($> 400$ Hz) 
mainly during the early phase in accretion, for short duration.

During the  $\dot M$ decline, accretion switches off or $\dot M$ remains very low for a discernible length of time.
Such behaviour was also found earlier by numerical computation \citep{podsiadlowski,main_jia&li}, and
it happens when magnetic braking is switched off ($\dot{J}_{\rm MB} = 0$) as the companion star becomes fully convective. 
During this period spin equilibrium is broken and $\nu$ does not change.
Eventually, accretion is switched on, and $\dot M$ decreases gradually.
In this phase, initially $r_{\rm co} < r_{\rm m} < r_{\rm lc}$, and hence the NS spins down in a propeller phase. Therefore, $r_{\rm co}$ increases and eventually becomes equal to $r_{\rm m}$. 
Thus, spin equilibrium is established again and $\nu$ tracks $\nu_{\rm eq}$ up to a relatively lower value at the end of the LMXB phase.
Note that we numerically find such a $\nu$-evolution trend for many other cases, although substantial variations from such a trend are also found (see later).
This shows that the long-term $\nu$-evolution is typically complex because the $\dot M$ evolution is complex.

The $\nu$-evolution is expected to happen via two different modes and be more complex, if the accretion is transient/episodic for certain periods of the LMXB phase \citep[see section~\ref{subsec: Mass Accretion}; ][]{sudip2021}.
Indeed, $\dot M < \dot{M}_{\rm crit}$ in the last part of the LMXB phase in Figure~\ref{Consistency Check}. 
We expect $\nu$ to approach an equilibrium spin frequency several times higher than $\nu_{\rm eq}$ during this period, and hence $\nu$ should increase \citep{sudip2021}.
However, in this paper, we study the $\nu$-evolution for non-episodic accretion, and will study more complex cases for episodic accretion using MESA in a subsequent work.

\begin{figure*}
\centering
\includegraphics[width = \columnwidth, scale = 1]{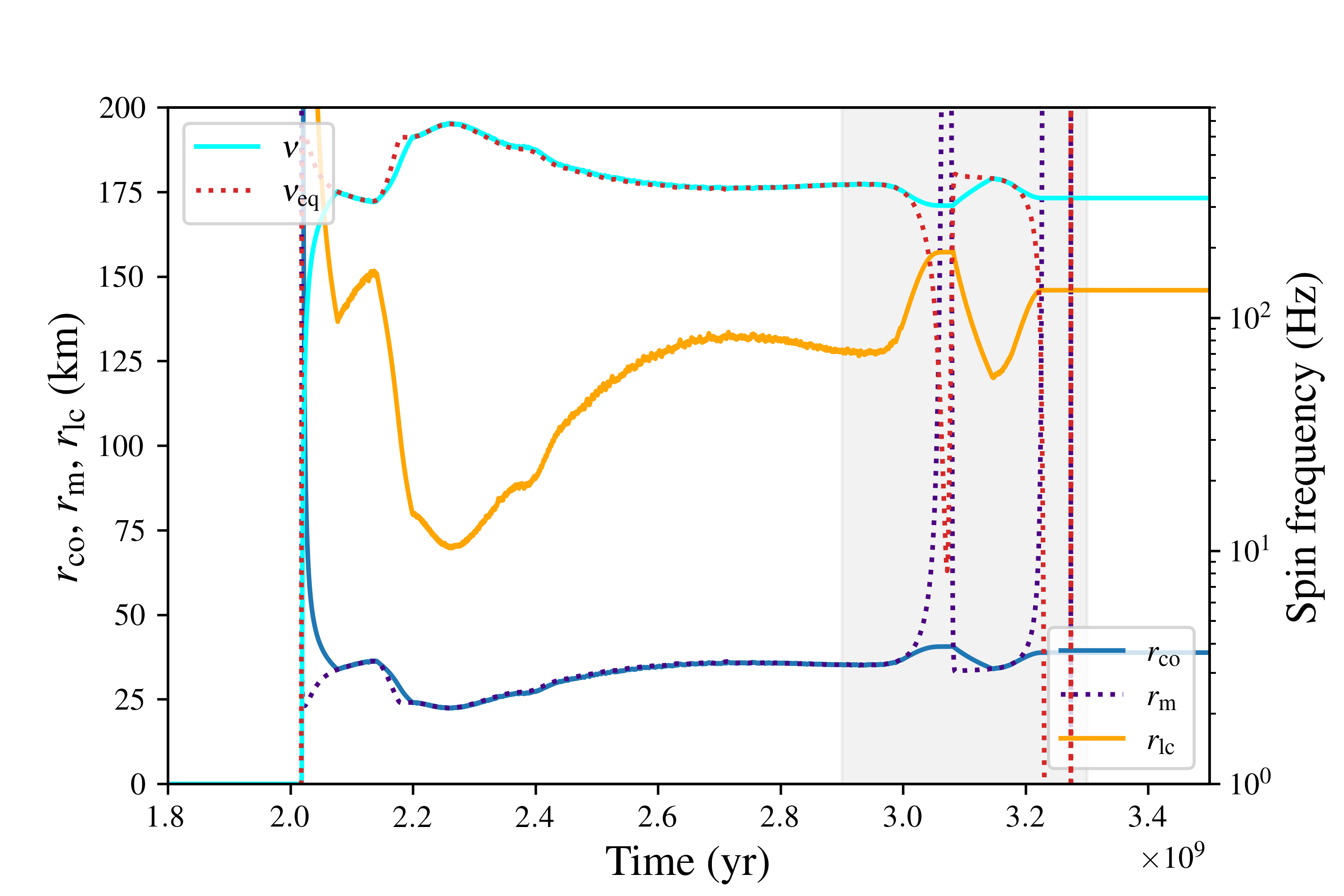}
\includegraphics[width = \columnwidth, scale = 1]{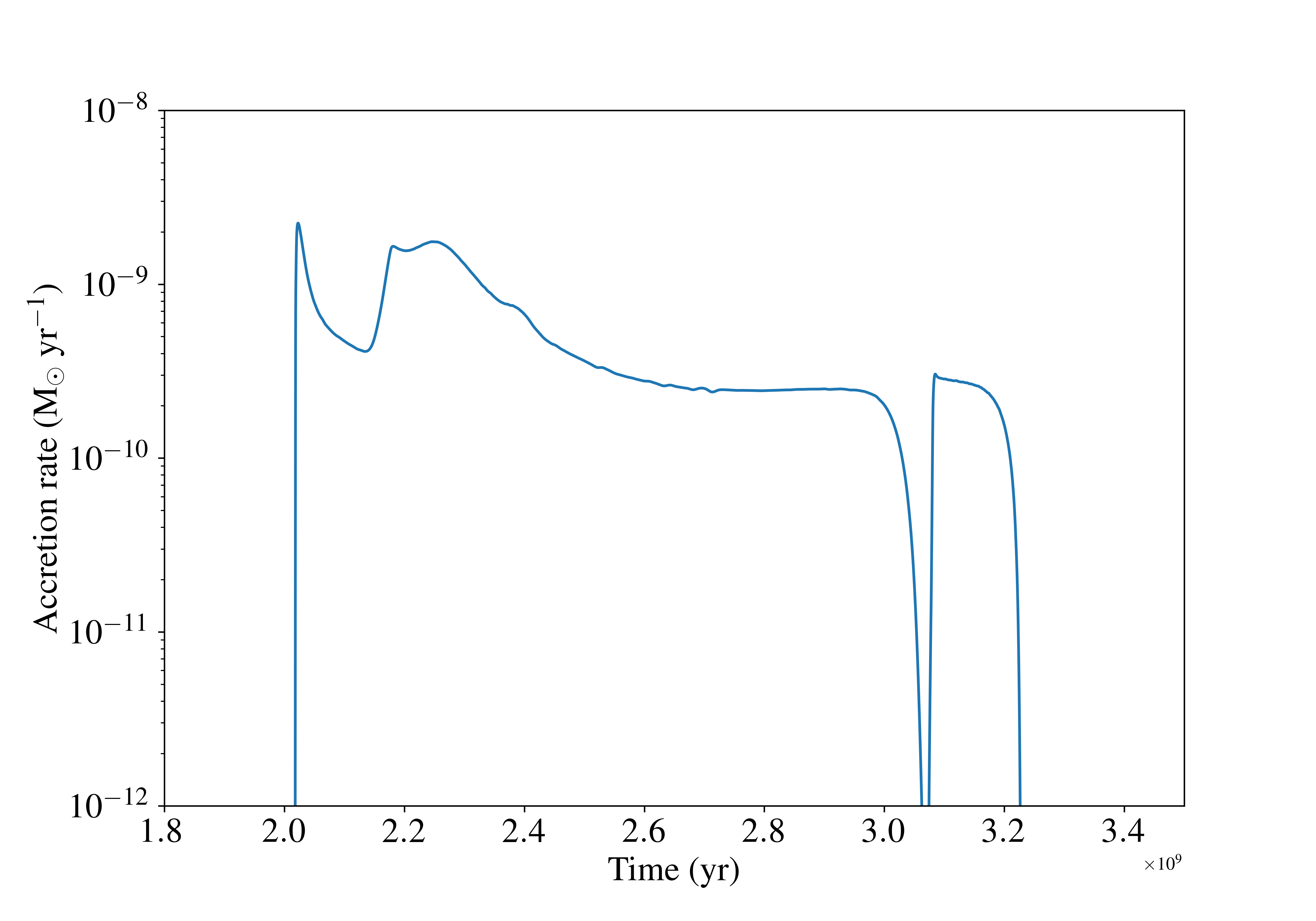}
\caption{Evolution of parameters (characteristic radii, spin frequencies, accretion rate $\dot{M}$) of a NS LMXB, as computed using MESA.
Here, we assume an initial orbital period 
 of
$1.0$ day and initial companion mass of 1.5 $\rm M_{\rm \odot}$. Other parameters are fixed at their canonical values (see Table~\ref{table1}).
These panels show the breaking from spin equilibrium [left panel] in the last phase of accretion [right panel] (see section~\ref{subsec: general example}).} 
\label{divergence}
\end{figure*}

In Figure~\ref{Consistency Check}, $\nu$ tracks $\nu_{\rm eq}$ till the end of the LMXB phase, and hence, given the substantial decrease of $\dot M$, the final $\nu$ value becomes low.
However, for certain parameter values, and if $\dot M$ declines relatively sharply in the last part of the LMXB phase, $\nu$ cannot sufficiently decrease given the relatively short time during which the negative torque operates in the propeller phase. Thus, $r_{\rm co}$ cannot track the fast increase of $r_{\rm m}$ and $\nu$ cannot track $\nu_{\rm eq}$. Thus, there is a breaking from spin equilibrium and $\nu$ remains relatively high at the end of the LMXB phase.
An example of this scenario is shown in Figure~\ref{divergence}, which shows the evolution of $\dot M$, $r_{\rm m}$, $r_{\rm co}$, $r_{\rm lc}$, $\nu$, and $\nu_{\rm eq}$ for an initial companion mass of 1.5 $ \rm M_{ \odot}$.

\subsection{Effects of specific parameters}
\label{subsec: effect on spin evolution result}
\subsubsection{Effects of different cases of magnetic braking (MB) }\label{sub : magnetic_braking}
\begin{figure*}
\centering
\includegraphics[width= \columnwidth, scale = 1]{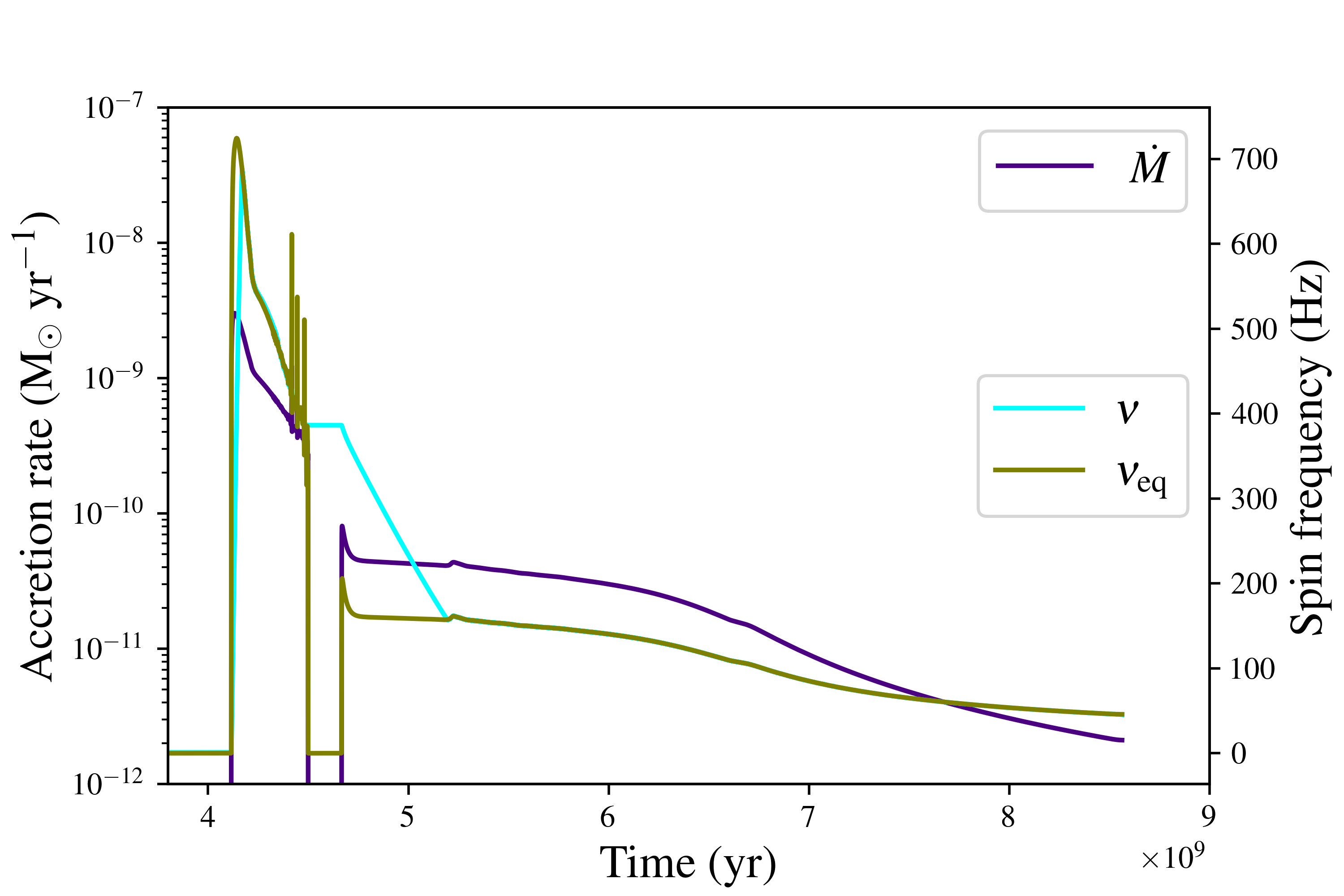}
\includegraphics[width = \columnwidth, scale = 1]{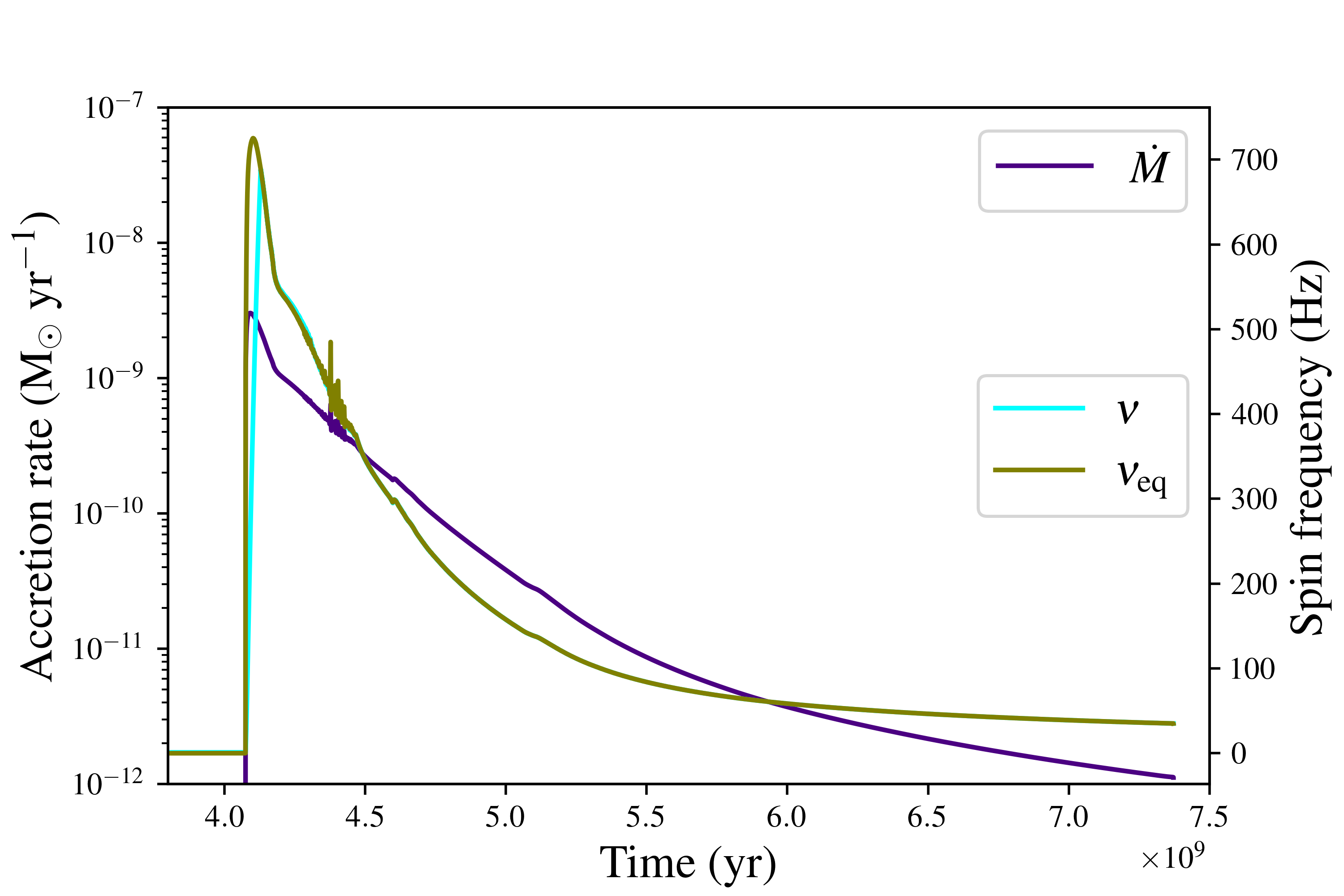}
\caption{Evolution of parameters (accretion rate $\dot{M}$, spin frequency $\nu$, and equilibrium spin frequency $\nu_{\rm eq}$) of a NS LMXB for two cases of magnetic braking (MB) prescription, as computed using MESA.
Here, we assume an initial orbital period 
 of
$1.7$ day and initial companion mass of 1.0 $\rm M_{\rm \odot}$.  
These panels show the effect of MB switched off when the companion star is fully convective [left panel] and MB always switched on [right panel] (see section \ref{sub : magnetic_braking}).}
\label{mag braking}
\end{figure*}

For our evolutionary computations, we generally switch off MB when the companion star becomes fully convective, following \citet{main_jia&li}, as a realistic option. 
This leads to a temporary drop in the mass transfer. 
On the other hand, MB could remain switched on
throughout the evolution \citep[see ][]{tailo}. 
Figure~\ref{mag braking} shows a comparison between these two scenarios.
With MB on throughout the evolution, unlike the first scenario, there is no temporary sharp drop in accretion, and thus $\dot M$ and $\nu$ decrease monotonically maintaining a spin equilibrium (Figure~\ref{mag braking}).

\begin{figure*}
\centering
\includegraphics[width = 85 mm, scale = 1]{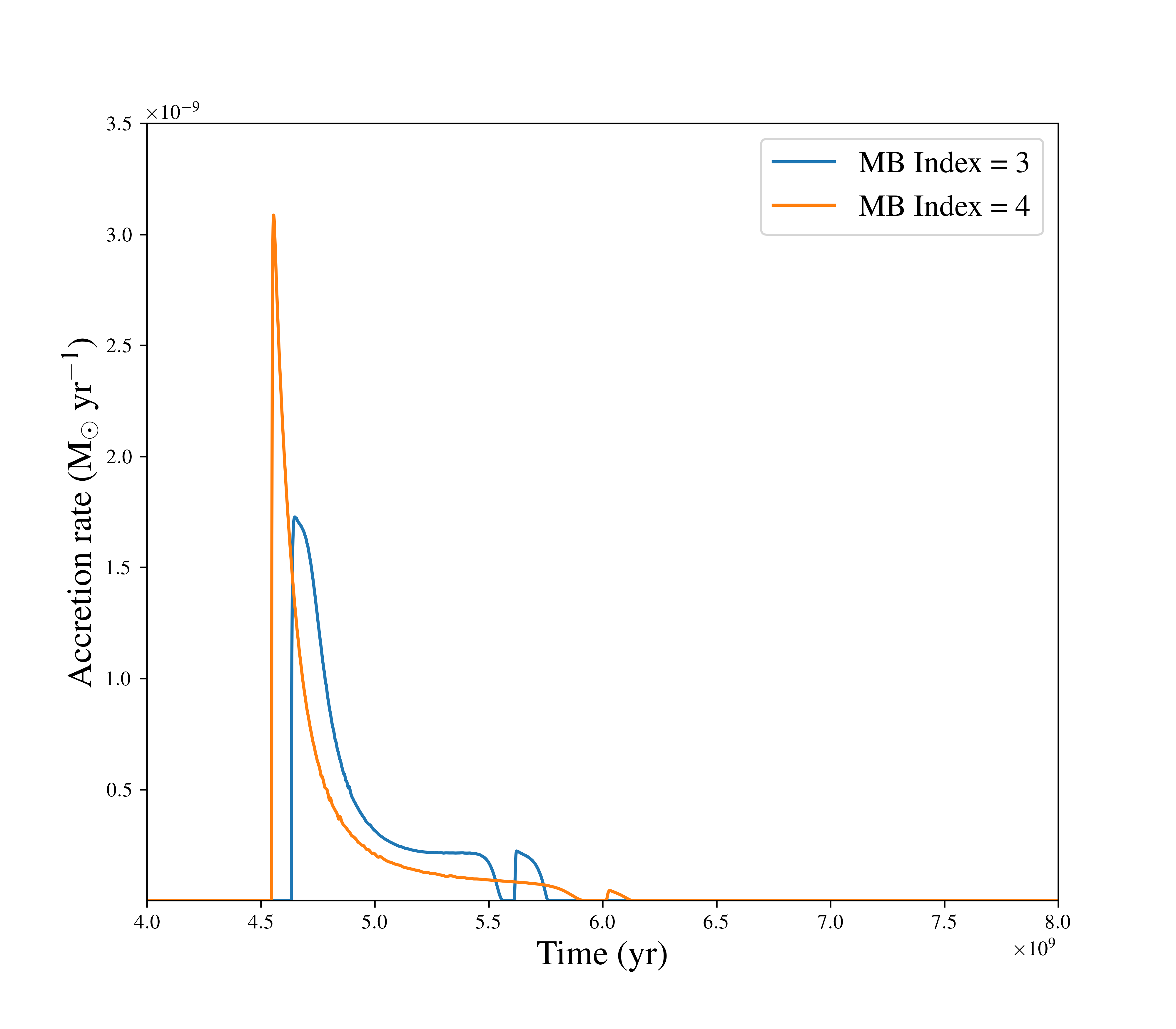}
\includegraphics[width = 85 mm, scale = 1]{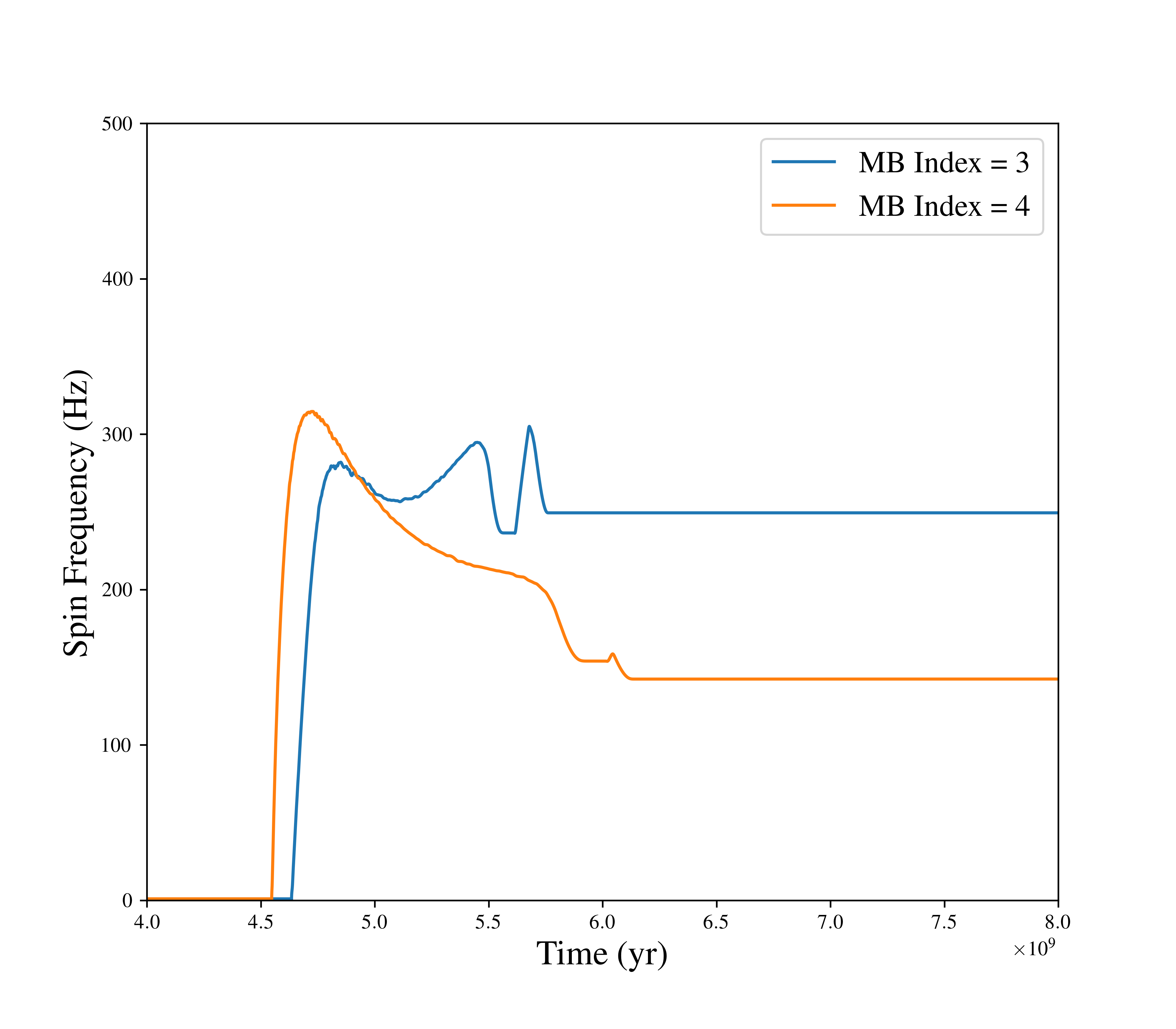}
\caption{Evolution of parameters (accretion rate [left panel] and NS spin frequency [right panel]) of an NS LMXB, as computed using MESA.
Here, we assume an initial orbital period 
 of
$1.5$ day and initial companion mass of 1.2 $\rm M_{\rm \odot}$. Other parameters are fixed at their canonical values (see Table~\ref{table1}).
This figure shows the effects of different magnetic braking (MB) index values on accretion rate and $\nu$ evolution (see section~\ref{sub : magnetic_braking}).} 
\label{MB index}
\end{figure*}

We also compare two prescribed magnetic braking (MB) index values: 3 and 4 (Figure~\ref{MB index}).
Note that a higher MB index implies a higher AML magnitude (see Equation~\ref{eq 5}), and hence a larger shrinking of the binary orbit.

 

\subsubsection{Effects of different values of NS mass quadrupole moment}
\label{subsec: Q}

\begin{figure*}
\centering
\includegraphics[width = \columnwidth, scale = 1]{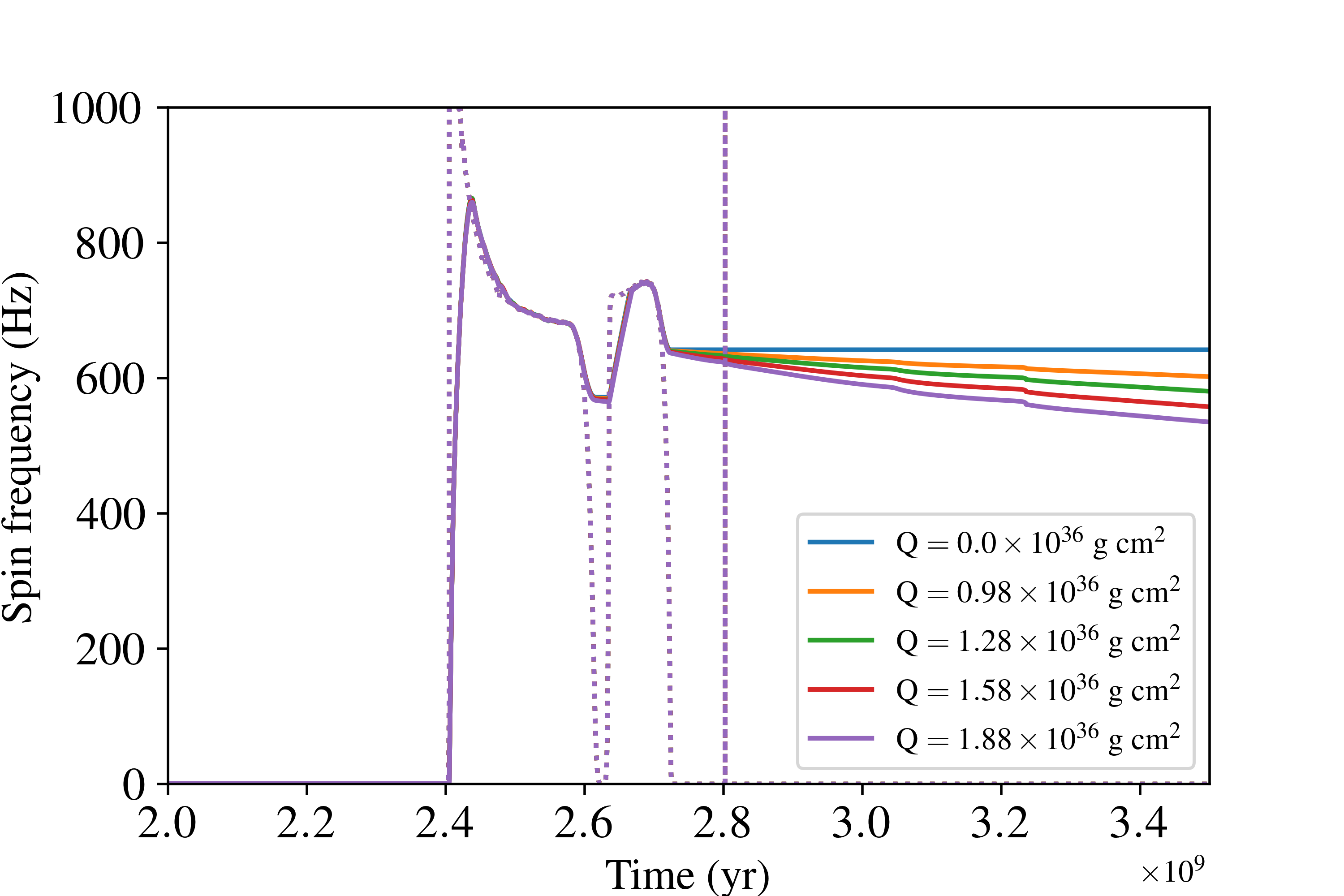}
\caption{Evolution of parameters (frequencies) of an NS LMXB for a range of NS mass quadrupole moment values, as computed using MESA.
Dotted line represents $\nu_{\rm eq}$, while solid line is for $\nu$.
Here, we assume an initial orbital period 
 of
$2.4$ days and initial companion mass of 1.5 $\rm M_{\rm \odot}$. Other parameters are fixed at their canonical values (see Table~\ref{table1}).
This figure shows that the relative effect of the gravitational torque $N_{\rm GW}$ on spin evolution is negligible during the accretion or LMXB phase, but is substantial when the accretion stops (see section~\ref{subsec: Q}).}
\label{Q case}
\end{figure*}

\begin{figure*}
\centering
\includegraphics[width= \columnwidth, scale = 3]{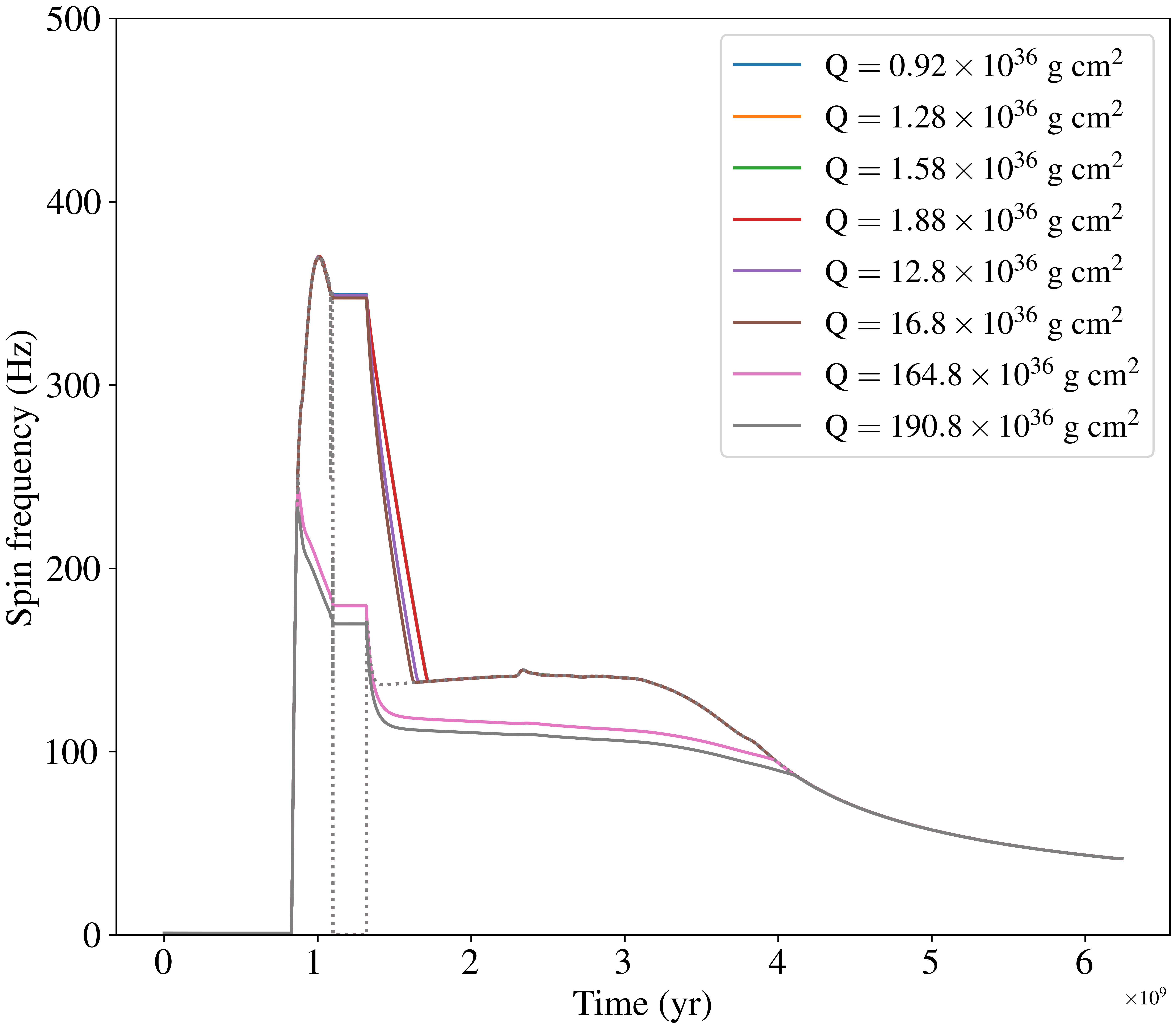}
\caption{Evolution of NS spin frequency ($\nu$) for wider range of Q values in LMXB phase. Dotted line represents $\nu_{\rm eq}$, while solid line is for $\nu$.  Here we assume initial companion mass of 1.0 $\rm M_{\rm \odot}$ and initial orbital period of 1 day, with other parameters being fixed at canonical values.  As shown in the plot, higher order Q values cause spin down of NS significantly, even during LMXB phase (see Section \ref{subsec: Q}).}
\label{Q_1_1}
\end{figure*}

If the NS has an asymmetric mass distribution around its spin axis, then it should spin down due to gravitational radiation and the corresponding torque ($N_{\rm GW}$; Equation~\ref{eq 24}).
In Figure~\ref{Q case}, we examine the effect of $N_{\rm GW}$ on $\nu$-evolution.
We find that for realistic values of the NS mass quadrupole moment $Q$ \citep{ellipcity_quadrupole}, the effect of $N_{\rm GW}$ is negligible compared to that of the torque due to accretion. However, after the LMXB phase, the effect of gravitational radiation on $\nu$-evolution becomes significant in a longer time scale.
As indicated in Figure~\ref{Q case}, this effect is significant when the $\nu$ value at the end of the LMXB phase is relatively high (e.g., by breaking from spin equilibrium).
This is expected because $N_{\rm GW} \propto \nu^5$ (Equation~\ref{eq 24}).
In Figure \ref{Q_1_1} we consider higher values of $Q$, corresponding to higher ellipticities of the NS of order of $10^{-8}$ and $10^{-7}$. Such higher values show a considerable effect in reducing the spin frequency even during the accretion phase. However, for individual MSPs, we perhaps  expect lower ellipticity values \citep[e.g., $\sim 10^{-9}$; ][]{Woan_2018,ellipcity_quadrupole}.


\subsubsection{Effects of NS magnetic field}
\label{subsec: mag}

\begin{figure*}
\centering
\includegraphics[width = \columnwidth, scale = 3.0]{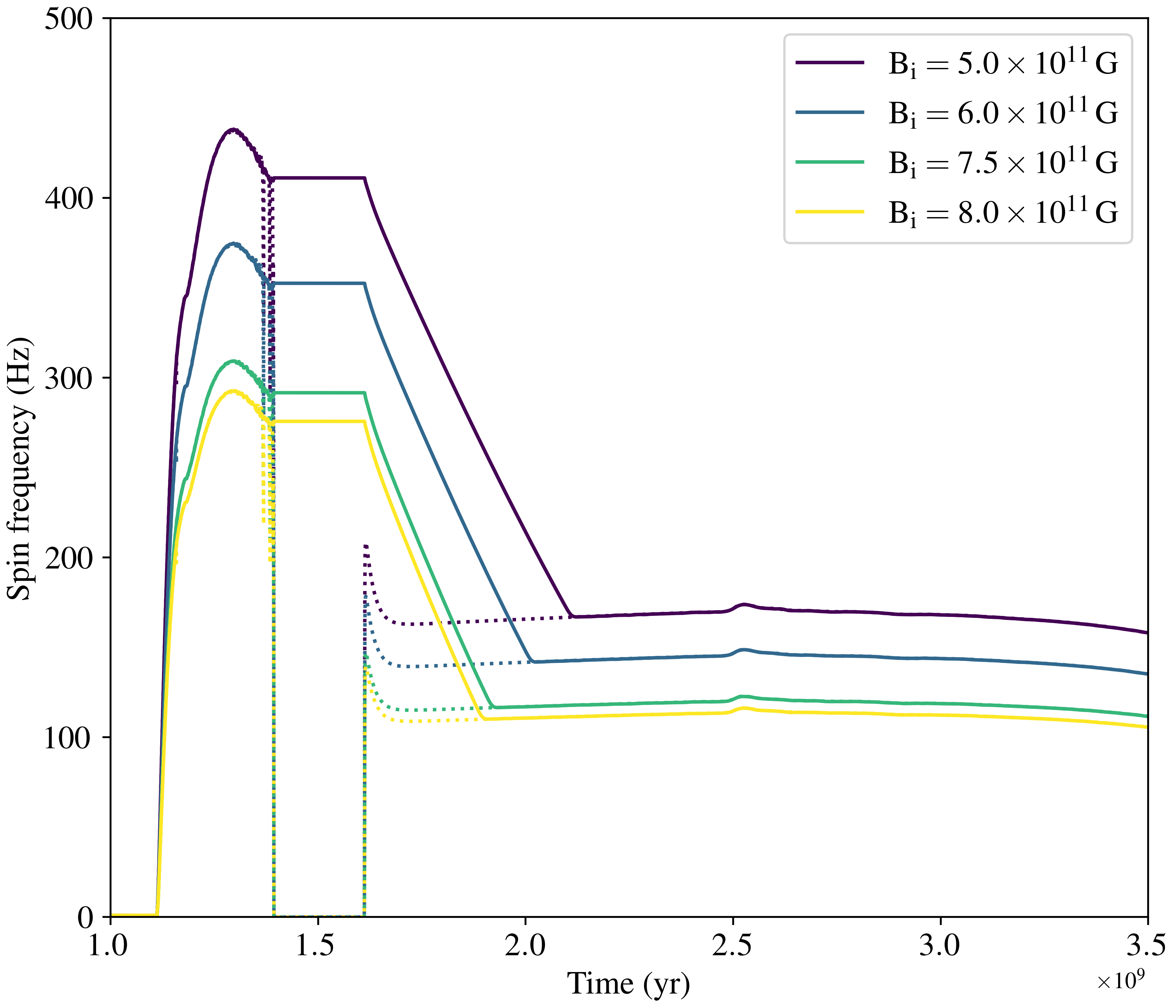}

\caption{Evolution of parameter (spin frequencies) of an NS LMXB for different values of initial NS magnetic field ($B_{\rm i}$), as computed using MESA (section~\ref{subsec: mag}). Corresponding to different initial magnetic values, the final values of magnetic field lie between $1.3\times 10^8$G - $3\times 10^8$G (Eq.\ref{B eq}).
Here, we assume an initial orbital period 
 of
$1.1$ day and initial companion mass of 1.0 $\rm M_{\rm \odot}$. Other parameters are fixed at their canonical values (see Table~\ref{table1}). 
Dotted line represents $\nu_{\rm eq}$ , while solid line is for $\nu$.}
\label{magnetic}
\end{figure*}

For a higher NS surface dipole magnetic field, $r_{\rm m}$ is higher and $\nu_{\rm eq}$ is lower (see Equations~\ref{eq 13} and \ref{eq 23}), and hence the NS attains lower $\nu$ values. This is shown in Figure~\ref{magnetic}. 
This figure shows that the qualitative nature of evolution of frequencies and characteristic radii do not depend on the $B$ value. However, we note that with evolving magnetic field (Eq \ref{B eq}), the maximum spin frequency decreases compared to that of a fixed NS magnetic field value.


\subsubsection{Effects of $\beta$}
\label{subsec:beta}
The fractional mass loss efficiency $\beta$ considerably affects the accretion rate and hence the NS evolution of the spin frequency (Figure~\ref{beta variation}).  
For low values of $\beta$ (e.g., $0.1$), $\dot M$ is high, and hence very high $\nu$ values ($\approx 1100$ Hz) are attained at the initial stage of the LMXB phase (Figure~\ref{beta variation}).
$\beta$ also affects the $M_{\rm NS}$ evolution significantly because a larger amount of mass is transferred to the NS for a lower $\beta$.
The effects of $\beta$ on $M_{\rm NS}$ and AML also slightly affect $P_{\rm orb}$ and LMXB duration (Figure~\ref{beta variation}). Overall, $\beta$ can play a significant role in matching the parameters of any accreting source, (see section \ref{sec:AMXPs}).

\subsubsection{Effect of irradiation}\label{subsec:irradiation}
\begin{figure*}
\centering
\includegraphics[width = \columnwidth, scale = 4.5]{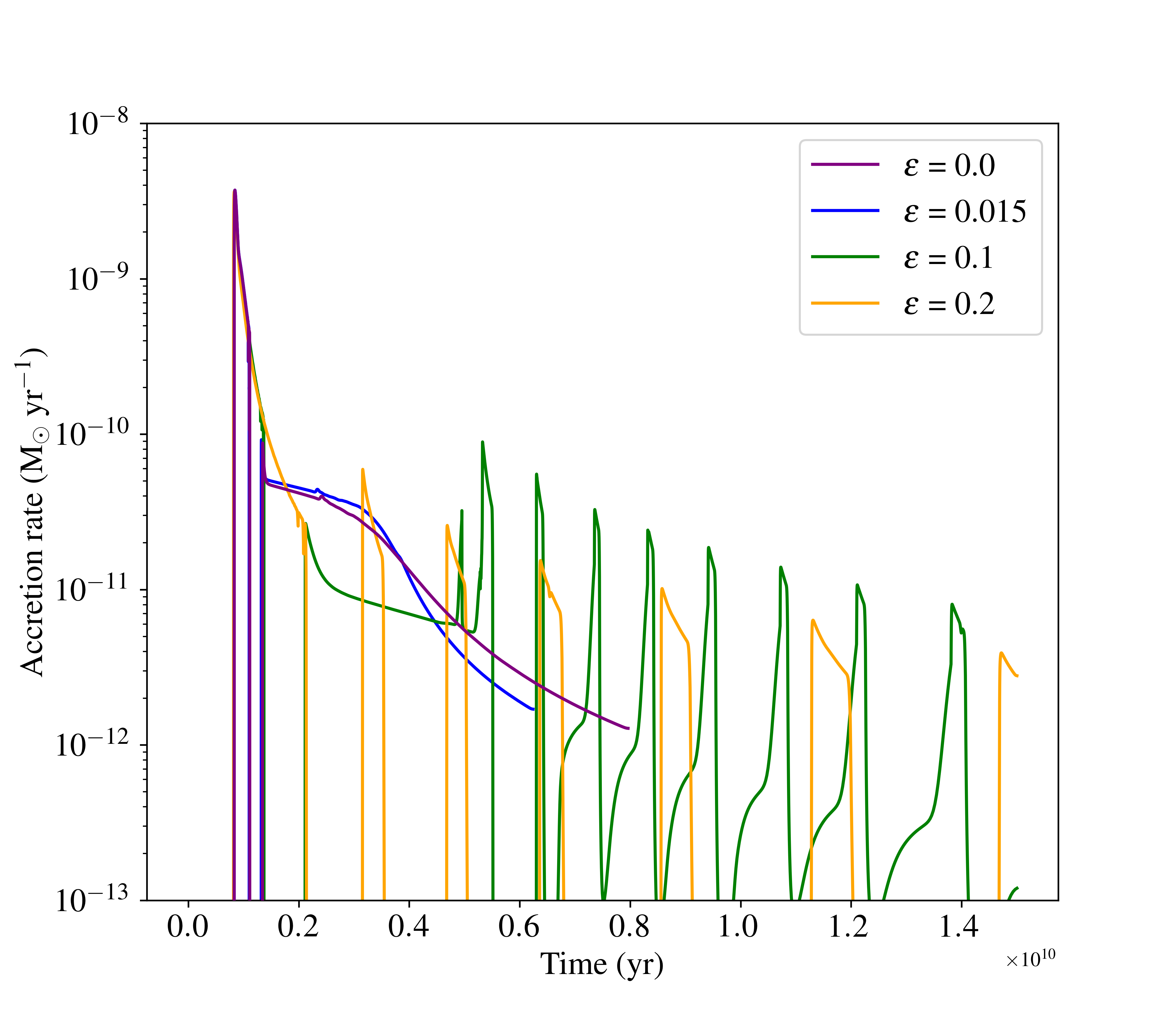}
\caption{This plot shows the effect of different irradiation efficiencies ($\epsilon$) on the accretion rate evolution of a NS LMXB, as computed using MESA. 
Here, we use 
parameters set at our canonical values (see Table~\ref{table1}). 
We take $\epsilon = 0.015$ as our default value for numerical computation. 
The accretion rate evolution is not cyclic for this low $\epsilon$ value, as well as for non-irradiation.  
However, higher irradiation  efficiencies ($\epsilon$ = 0.1, 0.2) show cyclic nature in accretion rate evolution}
\label{irradiation_eff}
\end{figure*}

We consider some cases with different irradiation efficiency and without irradiation, since it can affect the accretion rate ($\dot M)$ evolution. In Figure \ref{irradiation_eff}, it is shown that higher order irradiation efficiencies ($\epsilon = 0.1, 0.2$) cause the mass transfer rate ($\dot M$) to be cyclic in nature \citep[similar to the effect mentioned in][though initial parameters are different]{Benvenuto_2015,Lan_2023}. However, with our default model, which has relatively small value of $\epsilon $ = 0.015 \citep[following][]{goodwin}, this cyclic nature is not present.

\subsubsection{No significant effect of some other parameters}\label{other_parameters}

Some other parameters in Table \ref{table1} do not have a significant impact on the evolutionary tracks and values of other parameters. 
For example, variations in the proportionality constant ($\eta$) of the torque in the propeller phase, NS moment of inertia coefficient $A$ \citep[see][for a range]{sudip_newastron}, and different initial $\nu$ values do not show any significant effect on the $\nu$-evolution.


\subsection{Trends of NS LMXB parameter evolution}
\label{subsec: detailed-effect}

After probing the effects of specific parameters, now we systematically describe the trends of evolution of the following parameters: masses of both the binary components ($M_{\rm NS}$ and $M_{\rm comp}$),  accretion rate $\dot{M}$, binary orbital period $P_{\rm orb}$, three characteristic radii ($r_{\rm m}$, $r_{\rm co}$ and  $r_{\rm lc}$), the NS spin frequency $\nu$ and equilibrium spin frequency $\nu_{\rm eq}$.
We study the trends varying each of three parameters (initial $P_{\rm orb}$, initial $M_{\rm comp}$, and the fractional mass loss $\beta$), one at a time. 

In Figure~\ref{period days variation}, we show the computed evolution of the above-mentioned parameters for six values of initial $P_{\rm orb}$ in the range mentioned in Table~\ref{table1}. Other input parameter values are fixed at their canonical values (see
Table~\ref{table1}).
Figure~\ref{companion mass variation} shows the evolution of the same parameters, but for five values for initial $M_{\rm comp}$, keeping the other input parameter values fixed at their canonical values.
Similarly, Figure~\ref{beta variation} shows the effects of variation in $\beta$.


The following two subsections describe the cases with low initial $M_{\rm comp}$ and low initial $P_{\rm orb}$ (contracting systems) and those with high initial $M_{\rm comp}$ and high initial $P_{\rm orb}$ (widening systems). 

\subsubsection{Cases of low initial $M_{\rm comp}$ and low initial $P_{\rm orb}$}
\label{subsubsec : contracting}

These are usually contracting systems \citep[see][for the classification]{main_jia&li,istrate}, i.e., their $P_{\rm orb}$ overall decreases with the binary evolution (see the bottom-right panel of Figure~\ref{companion mass variation}). 
Most of the cases within our used range of  $P_{\rm orb}$ for initial $M_{\rm comp} = 1 \rm M_{\rm \odot}$ are contracting ones, but for fixed initial $P_{\rm orb} = 1$ day, those with initial $M_{\rm comp}$ 
greater than 1.2 $\rm M_{\rm \odot}$ show widening behaviour. 

\begin{figure*}
\centering

\includegraphics[width= \columnwidth, scale = 1]{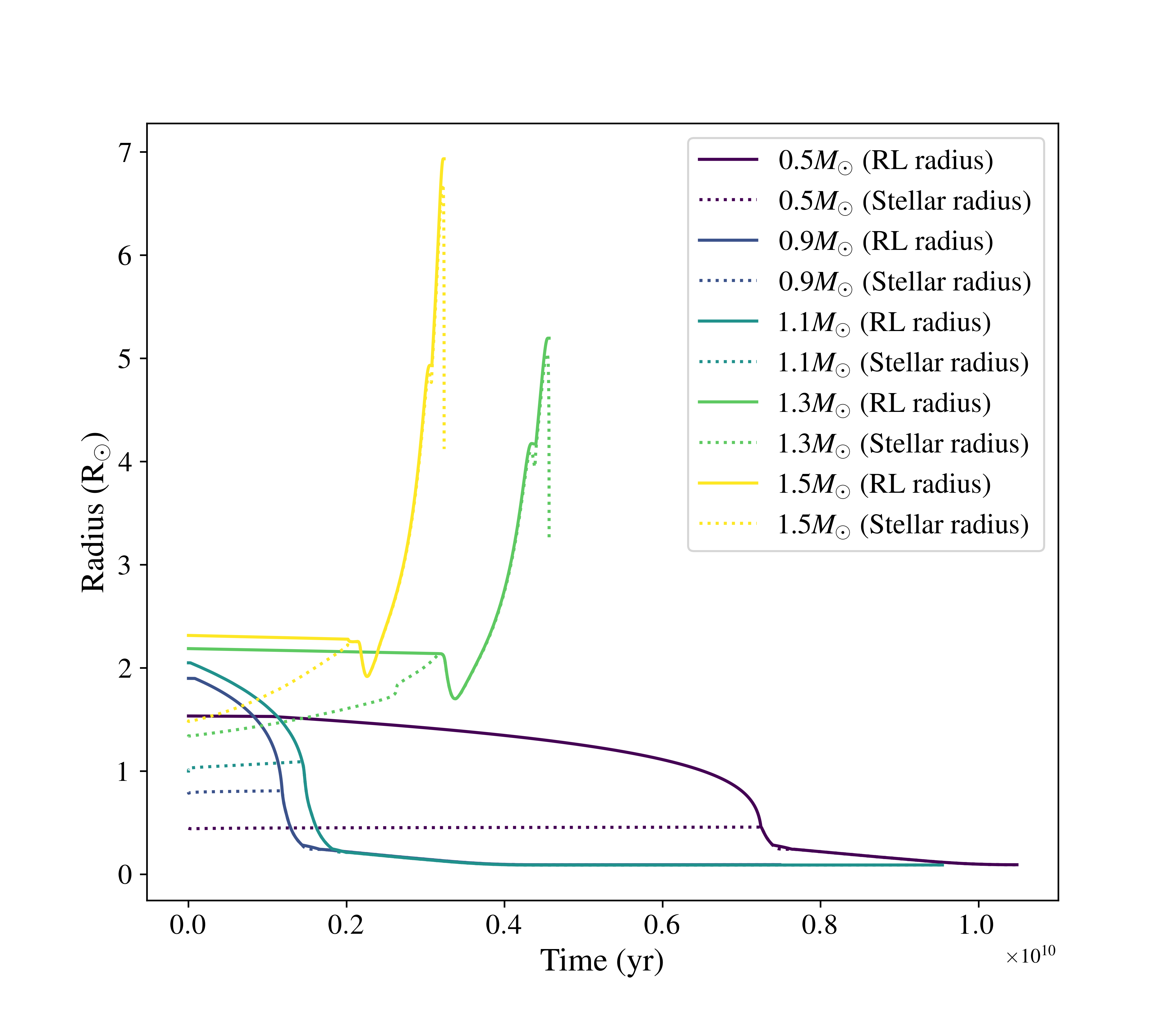}
\caption{Evolution of companion star Roche lobe radius, stellar radius for different initial values of companion star mass. Other initial parameters are fixed at their canonical values (see Table~\ref{table1}). The plot helps in explaining trends of RLOF onset times for LMXB systems (see section \ref{subsubsec : contracting}).}
\label{companion radius}
\end{figure*}

For the fixed initial $P_{\rm orb} = 1$ day,
the age at which the mass transfer begins, i.e., the Roche lobe overflow (RLOF) onset age depends on the initial $M_{\rm comp}$ values in a complex way
(see Figure~\ref{companion mass variation}). 
Note that this age is measured from the time of assumed initial values for a system. 
Figure~\ref{companion mass variation} shows that the RLOF onset age decreases from initial $M_{\rm comp} = 0.5 \rm M_{\rm \odot}$ to initial $M_{\rm comp} = 0.9 \rm M_{\rm \odot}$, increases up to initial $M_{\rm comp} = 1.3 \rm M_{\rm \odot}$ through $1.1 \rm M_{\rm \odot}$, and again decreases for initial $M_{\rm comp} = 1.5 \rm M_{\rm \odot}$.
This happens because of various competing effects before the RLOF onset time, which can be seen from Figure~\ref{companion radius}.
Here, we explain these effects.
(i) Companion's Roche lobe (RL) radius decreases (as the orbit shrinks due to a negative $\dot J$ primarily because of the magnetic braking) and the companion radius remains almost the same for relatively lower mass stars 
and this is how such a companion fills its Roche lobe (initial $M_{\rm comp} = 0.5 \rm M_{\rm \odot}, 0.9 \rm M_{\rm \odot}, 1.1 \rm M_{\rm \odot}$ in Figs.~\ref{companion mass variation} and \ref{companion radius}).
(ii) For relatively higher mass companions (initial $M_{\rm comp} = 1.3 \rm M_{\rm \odot}, 1.5 \rm M_{\rm \odot}$ in Figs.~\ref{companion mass variation} and \ref{companion radius}), 
the companion radius increases significantly as the star transforms to become a red giant and this is how such a companion fills its Roche lobe.
Thus, for relatively lower mass stars, the RLOF age is determined by the competition among the companion radius and RL radius (which are higher for higher mass) and the rate of RL radius decrease (see Figure~\ref{companion radius}). As Fig.~\ref{companion radius} clearly shows, this competition causes the non-monotonic behaviour of the RLOF onset age with the initial companion mass.
But for relatively higher mass companions, as a more massive star evolves, fills its Roche lobe and becomes a red giant faster, the RLOF onset age is smaller for a more massive star (see Figs.~\ref{companion mass variation}, \ref{companion radius}).

However, for systems with fixed initial $M_{\rm comp}$, the age at which the mass transfer begins increases with the increase in initial $P_{\rm orb}$ throughout the used range, as shown in Figure~\ref{period days variation}. 
The maximum value of spin frequency attained by the NS is higher for contracting systems (Figure~\ref{period days variation}). As explained earlier, the mass transfer stops for short time, when the magnetic braking is switched off due to the companion star becoming fully convective, leading to no change in $\nu$ temporarily, as shown in Figure~\ref{Consistency Check}. This trend is less conspicuous for 
cases of higher companion star mass, which implies that such companion stars do not usually become fully convective in the LMXB phase (Figure~\ref{companion mass variation}).

In the systems with initial $M_{\rm comp} = 1 \rm M_{\odot}$ and initial $P_{\rm orb} \leq 1$ day, $\dot{M}$ remains higher than $\dot{M}_{\rm crit}$ for our considered parameter values, making them persistent accretors for a major fraction of the LMXB phase. 
It is observed that the binary evolution continues for longer durations and the final $\nu$ values are lower than 100 Hz for low initial $M_{\rm comp}$ and low initial $P_{\rm orb}$ cases. 
This is somewhat different from the observed spin frequencies of MSPs. 

In a few of these systems, such as initial $M_{\rm comp} = 0.9 \rm M_{\rm \odot}$ and 1.1$\rm M_{\rm \odot}$ in Figure~\ref{companion mass variation},  $P_{\rm orb}$ starts evolving earlier than the age at which accretion starts significantly. This is because of AML due to magnetic braking and gravitational radiation (see section~\ref{section_AML}).

\subsubsection{Cases of high initial $M_{\rm comp}$ and high initial $P_{\rm orb}$}
\label{subsubsec: widening}

These are usually widening systems, i.e., their $P_{\rm orb}$ overall increases with the binary evolution \citep[see][for the classification]{main_jia&li}, as shown in the bottom-right panel of Figure~\ref{companion mass variation}. 
We find a few such cases for initial $P_{\rm orb} > 2$ days,  and initial $M_{\rm comp} = 1 \rm M_{\rm \odot}$.  
But, for initial $M_{\rm comp} = 1.5 \rm M_{\rm \odot}$, the cases with initial $P_{\rm orb} \gsim 0.7$ day show widening. 

In some of the widening systems \citep[with initial $M_{\rm comp} = 1.0 \rm M_{\odot}$ and initial $P_{\rm orb} > 2$ days; see][for discussion on bifurcation period]{bifurcation}, the final $\nu$ values are typically higher ($\nu \gsim 300$ Hz at the end the LMXB phase) than those for contracting systems, as shown in Figure~\ref{period days variation}. 
These cases also show shorter LMXB phases, and sometimes broken spin equilibrium.
The widening systems might also be transient (due to $\dot M < \dot{M}_{\rm crit}$) for most of their LMXB phase.

\section{General evolution of observed AMXPs}
\label{sec:AMXPs}



After discussing the general trends of evolution of LMXB parameters, we focus on finding cases which can reproduce the observed AMXPs during the LMXB phase.  
Most of the observed AMXPs \citep[see Table 1]{Salvo2022}, have $P_{\rm orb}$ of a few hours and a significant spin frequency in the range of 160 - 600 Hz, with the companion star of different types. In this context, considering all the set of parameters including that of NS, donor star and the orbital ones are necessary for predicting evolutionary models of AMXPs and matching the observational values. Previously, some authors \citep{xianghe,amxp_suvorov} looked into the evolution of some parameters of AMXPs. While these papers explored parameter space for orbital period, donor mass and its type \citep[in particular]{xianghe}, spin frequency evolution was not considered at all.  
In this section, we discuss possible evolutionary models for observed AMXPs including their spin frequency evolution, for the first time to the best of our knowledge.

\begin{figure*}
\centering
\includegraphics[width= \columnwidth, scale = 1]{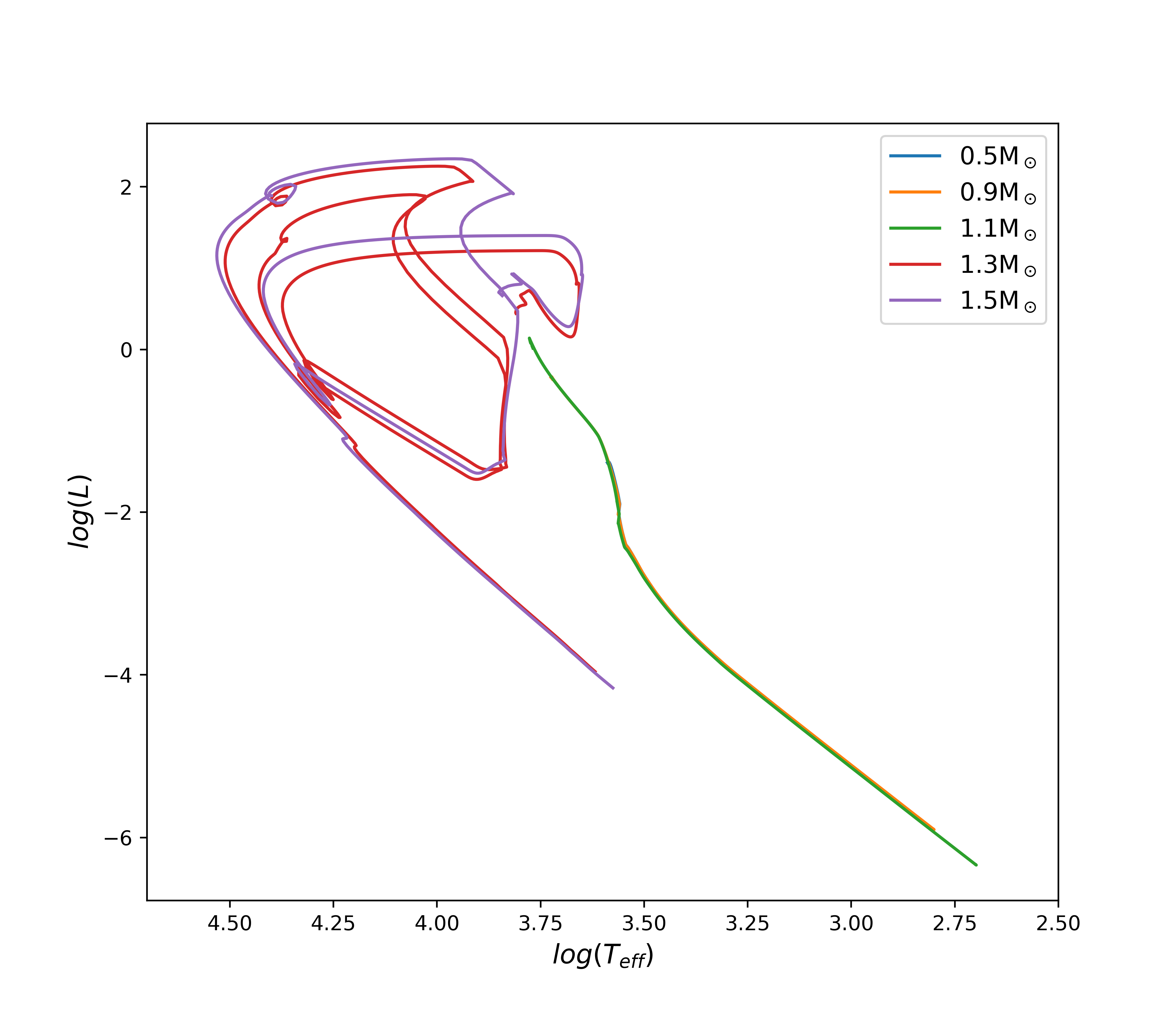}
\includegraphics[width= \columnwidth, scale = 1]{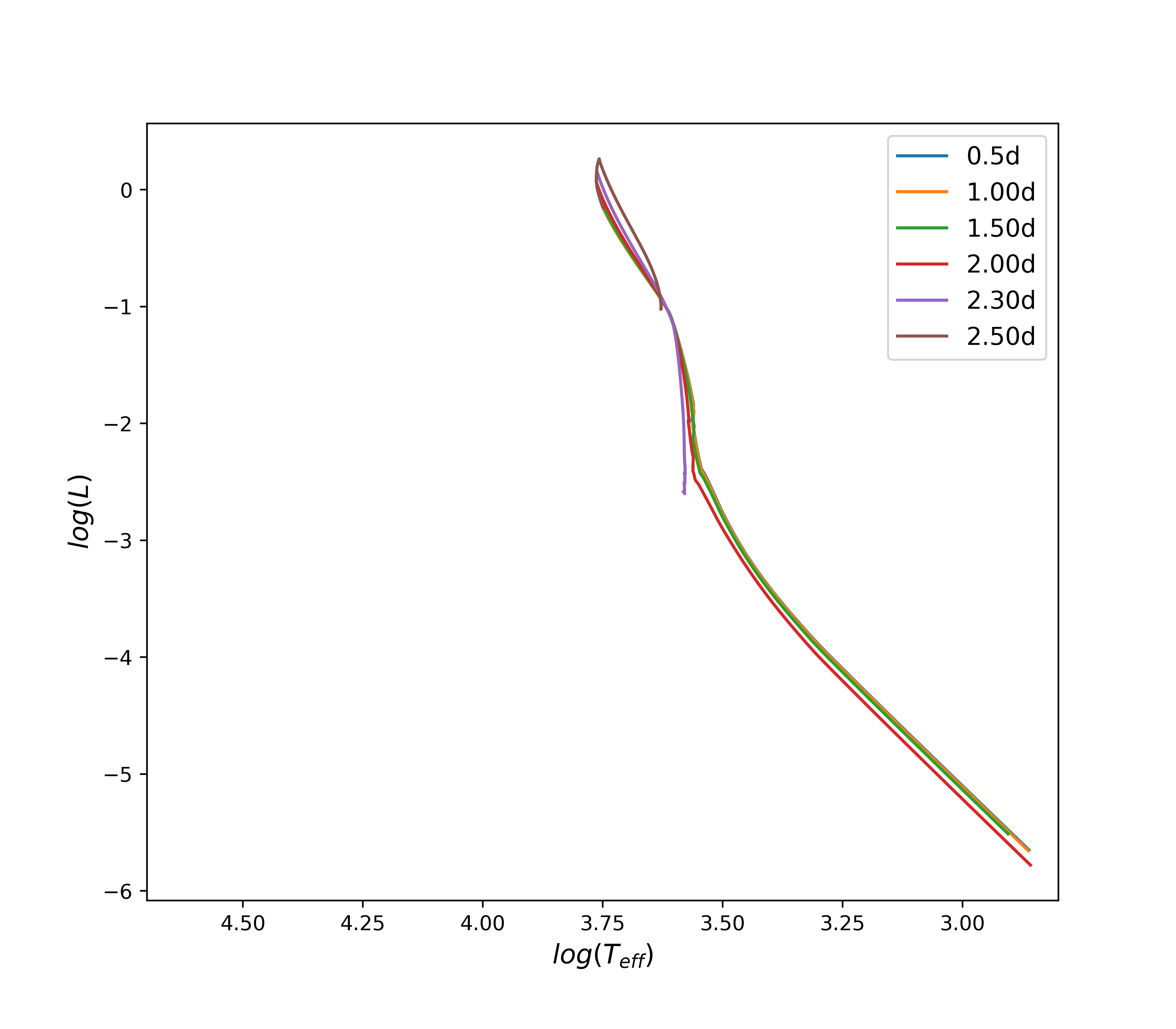}
\caption{HR diagram for companion stars, as calculated by MESA. The left plot shows the HR diagram of donor star at different values of its initial mass. The right plot shows HR diagram of donor star at six different initial orbital period values. All other initial parameters are fixed at their canonical values. The plots are useful to determine companion star type throughout the evolution, as discussed in sections \ref{subsubsec : contracting}and  \ref{sec:AMXPs}.}
\label{hr diagram}
\end{figure*}

For the initial models, we expect the convergent LMXB cases from our general results to produce the observed small orbital period of AMXPs in their accretion phase, unlike the divergent cases.
The evolution of companion stars from three categories: Main Sequence (MS), Brown Dwarf (BD), White Dwarf (WD)  play a crucial role too in predicting the models.
For instance, Table 1 of \citet{Salvo2022} shows that the systems with MS companions have considerably high $\nu$, low $P_{\rm orb}$ and also higher threshold for minimum companion star mass compared to other cases. 
LMXB cases with different initial $P_{\rm orb}$ for a fixed initial companion mass $\leq 1.2 \rm M_{\rm \odot}$, undergo RLOF during their main sequence phase only (see Figure \ref{hr diagram}). So, these cases can be explored as probable theoretical models for AMXPs with MS companions, exhibiting low $P_{\rm orb}$ and high $\nu$. Similarly, most of the AMXPs with WD companions have comparatively lower $\nu$ along with $P_{\rm orb} \approx 1~ \rm hr$. With the companion star already in the WD stage, these systems are estimated to be in their later phase of evolution. Among all the observed AMXPs, those with BD companions are challenging to reproduce with all their parameters including spin frequency, as these systems often have a high spin frequency at the extremely end phase of LMXB evolution.

\begin{figure*}
\centering
\includegraphics[width= 90 mm
, scale = 4]{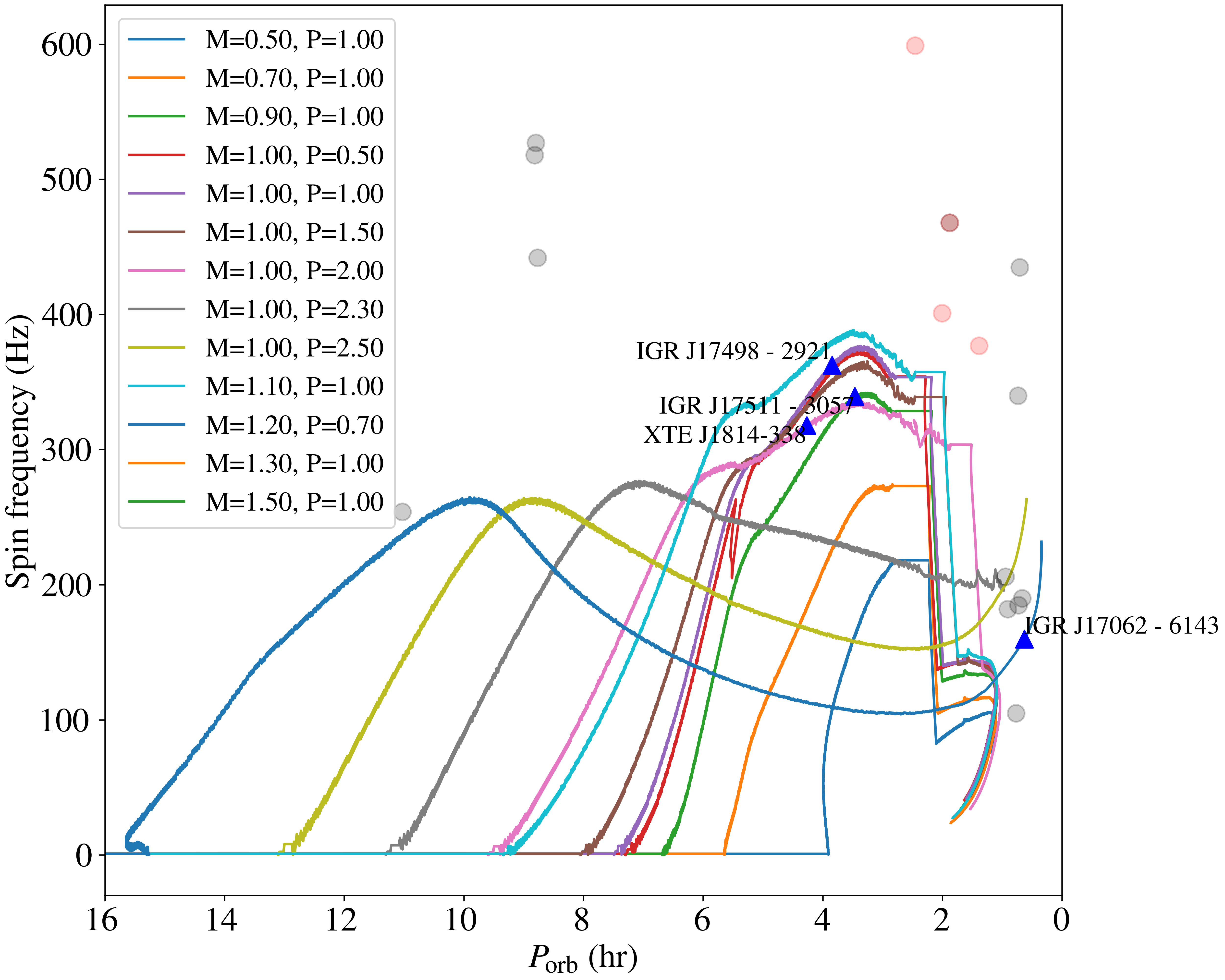}
\caption{Evolution of NS spin frequency ($\nu$) vs $P_{\rm orb}$ for various initial values of donor mass and orbital period. All other parameters are fixed at their canonical values. We indicate the NS spin frequency and orbital period for observed AMXPs in this plot. The blue triangles indicate the sources for which we exactly match the observed parameters. The grey dots indicate other observed AMXPs, for which we only indicate the possible parameter space. The red dots correspond to the AMXPs with Brown Dwarfs (BD) as companions. The plot also shows possible evolutionary models for various AMXPs along with the constraints, as discussed in section \ref{sec:AMXPs}.}
\label{nu vs porb}
\end{figure*}

\begin{figure*}
\centering
\includegraphics[width= 90 mm, scale = 4]{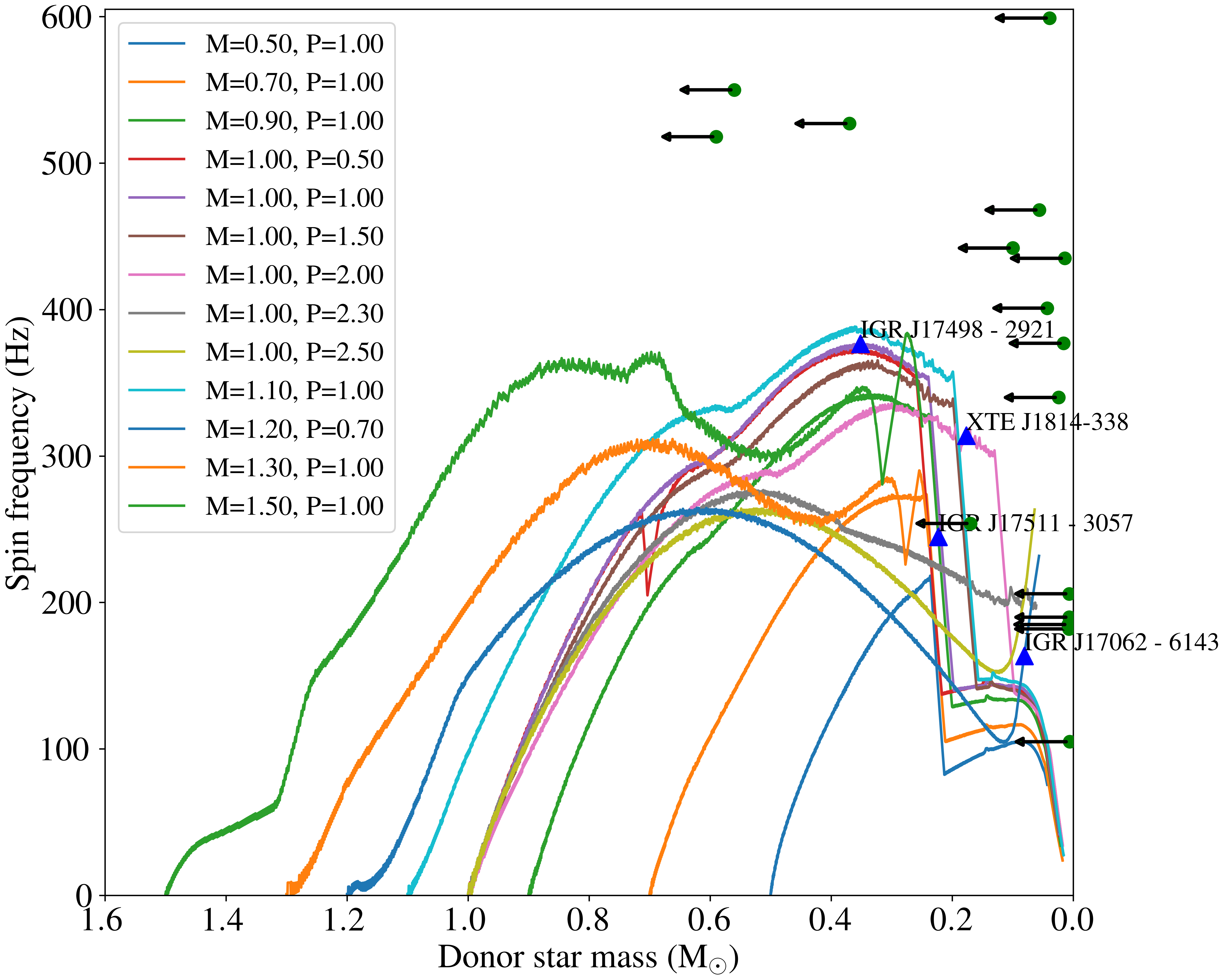}
\caption{Evolution of NS spin frequency ($\nu$) vs Donor mass ($\rm M_{\odot}$) for various initial values of donor mass and orbital period. All other parameters are fixed at their canonical values. We plot the AMXPs with minimum companion mass as green point and indicate the range above it with the arrow, as the donor star can take values greater than the minimum companion mass. Similar to Figure \ref{nu vs porb}, the blue triangles indicate the AMXP sources for which we exactly match the observed parameters. The plot shows interplay between donor star mass and the NS spin frequency as discussed in section \ref{sec:AMXPs}.}
\label{nu vs donor}
\end{figure*}

As shown in Figure \ref{magnetic} and \ref{beta variation} the initial NS magnetic field values and $\beta$ values affect the spin frequency of the pulsar. 
We pick one of our LMXB models which lies closest to the source and adjust the NS magnetic field within the estimated range for observed cases \citep{amxpmagDB} to obtain the spin frequency and $P_{\rm orb}$ within $1\%$ of the observed value for the same evolution timesteps. 
The distribution of AMXPs in the possible parameters space, with exactly matching the parameters of a few sources are shown the Figures \ref{nu vs porb} and \ref{nu vs donor}. We now discuss a few AMXPs for which we predict evolutionary models in detail and closely match with observational values.
\subsection{IGR J17511-3057}
\label{IGR J17511-3057}
This is an AMXP source with a relatively lower spin frequency of 245 Hz and orbital period $3.47$ hr with an MS companion star having minimum companion mass of 0.13 $\rm M_{\odot}$ \citep{IGR17511-2011, IGR17511-2016}. The rough estimate of the maximum magnetic field is $3.5\times 10^8$ G and the $\dot{\nu} = 1.45\times 10^{-13} \rm $  Hz $\rm s^{-1} $ \citep{IGR17511-2016}. To match this set of parameters, we need a convergent case with a low enough orbital period. The closest cases are those with initial companion mass 0.9$\rm M_{\rm \odot}$ and 1.1 $\rm M_{\rm \odot}$ for the fixed initial orbital period (see Figure \ref{companion mass variation}). For 0.9 $\rm M_{\rm \odot}$, we get an orbital period of 3.473 hr during the LMXB/accretion phase at an age of $1.337\times 10^9 $ years. Adjusting the initial NS magnetic field, taking $B_{\rm init} \sim 8\times 10^{11}$G produces a spin frequency of 245.02 Hz and $B = 3.59\times10^8$ G at the corresponding timestep, which falls within the estimated range \citep{amxpmagDB}. This best-fit model also shows that the companion star lies in MS phase (Figure \ref{hr diagram}) at that age, while its mass is 0.37 $\rm M_{\rm \odot}$, which matches with the observed limit. We are also able to closely match $\dot{\nu}$ during the accretion phase, which is $1.57\times10^{-13} \rm $ Hz $ \rm s^{-1}$ in our best-fit model. All the reproduced values are within $1\%$ error bar of the observed ones.

\subsection{XTE J1814-338}
\label{XTE J1814-338}
This AMXP has observed spin frequency of 314 Hz, the observed orbital period of 4.27 hr, minimum companion mass of 0.17 $\rm M_{\rm \odot}$ \citep{XTEJ1814}. Like the previous source, this also shows a low orbital period in the accretion phase and a considerably high spin frequency. We find our best-fit model for this source in one of the convergent cases with initial orbital period of 2.0 days and an initial companion mass being 1.0 $\rm M_{\rm \odot}$. We get the closest parameter values for the evolutionary timesteps at age $6.65\times 10^9$ years. At that age, orbital period comes out to be 4.263 hr, along with the spin frequency being 312.9 Hz. By adjusting the initial NS magnetic field, we get $B = 2.42\times 10^8$G which falls under the inferred magnetic field for the source \citep{amxpmagDB}, 
We also see that the companion star is in main sequence phase at that age (Figure \ref{hr diagram}). Thus from our general evolutions, parameter space for XTE J1814-338 can be obtained with reasonable accruacy.

\subsection{IGR J17062-6143}
\label{IGR J17062-6143}
This AMXP has a He WD companion star and its orbital period is 0.63 hr, spin frequency of 163.65 Hz \citep{Bult_2021_IGR17062,Strohmayer_2017_IGR17062}. Its spin frequency is less, compared to previous other cases and it has minimum companion mass limit of 0.006 $\rm M_{\rm \odot}$. This hints towards the fact that this system is in late phase of evolution when the companion star has reached the WD stage. For this kind of scenario, we consider initial companion star mass around 1.2 $\rm M_{\rm \odot}$ and a small initial orbital period of 0.7 day, so that the companion star can become WD within the LMXB phase itself, and orbital period remain convergent. We set our initial B value such that the decayed magnetic field at this age is $1.46\times10^8$ \rm G or $\mu = 1.66\times10^{26} \rm $ G $\rm cm^3$ \citep{Strohmayer_2017_IGR17062}. Our best fit model produces 0.629 hr orbital period and spin frequency of 163.9 Hz at an age of $7.634\times10^9$ yr. Its observed spin up rate is measured to be $3.77\times10^{-15} \rm$ Hz $\rm s^{-1}$, which also matches with our $\dot \nu = 1.7\times 10^{-15} \rm$ Hz $\rm s^{-1}$, in its order.

\subsection{IGR J17498-2921}
\label{IGR J17498-2921}
This system has a pulsar with spin frequency 401 Hz, observed orbital period of 3.84 hr. The companion is a main sequence star (MS), with an estimated minimum mass of 0.17 $\rm M_{\rm \odot}$ \citep{papitto_IGRJ17498}. We get our best fit model for this in initial parameters : 1.0 $M_{\rm \odot}$ donor mass and 1.0 period day. However, with the general evolution results (with $\beta = 0.5$ of this model, Figure \ref{period days variation}) we don't get spin frequency beyond 400 Hz. To adjust for the higher spin frequency, we take the $\beta$ to be 0.3, and initial B value of $9.5\times10^{11}$G. At age $9.63\times10^8$ yr from the model, we get an orbital period of 3.848 hr and spin frequency of 402.3 Hz. Magnetic field value at that age is $2.9\times10^8$ G, which falls within the range of estimates of it \citep{amxpmagDB}.\\

Apart from these exact matches, we indicate parameter space for other sources in Figures \ref{nu vs porb}, \ref{nu vs donor}. For eg. IGR J17591-2342 has spin frequency of 527 Hz and $P_{\rm orb}$ of 8.80 hr. To get such a high spin frequency we use a model with less $\beta$ values and lower B values ( Figure \ref{magnetic}, \ref{beta variation}). The relatively higher minimum companion mass also denotes the early phase of evolution for this system. Similarly for other sources having high spin frequency (> 400 Hz), we can get the orbital period and spin frequency parameter space from convergent systems with adjusted $\beta$ and B values. The companion star type which is MS, also matches for these sources. Similarly for systems with WDs, we can have higher initial companion mass systems which can evolve to a WD, but is convergent with orbital period, given their very small observed orbital period. As discussed above, to get the exact match for other sources, we can run more dense grid around the initial values from the parameter space indicated in general evolutions.
However, we can’t predict the initial parameters for one particular class of AMXPs with BD companion. These AMXPs have high spin frequency in the range of 400 - 500 Hz. In our case, at the end of evolution with companion mass $\leq 1.1 \rm M_{\rm \odot}$, we get BDs but spin frequency doesn’t remain very high at that time. We can take higher initial orbital period ($\geq2.3$ days) to get high spin frequency at the end, but those cases simultaneously don't produce small orbital period observed in these systems. 

\section{Discussion}
\label{sec:discussion}

\begin{figure*}
  \sbox0{\begin{tabular}{@{}cc@{}}
    \includegraphics[width=170mm, scale = 0.8]{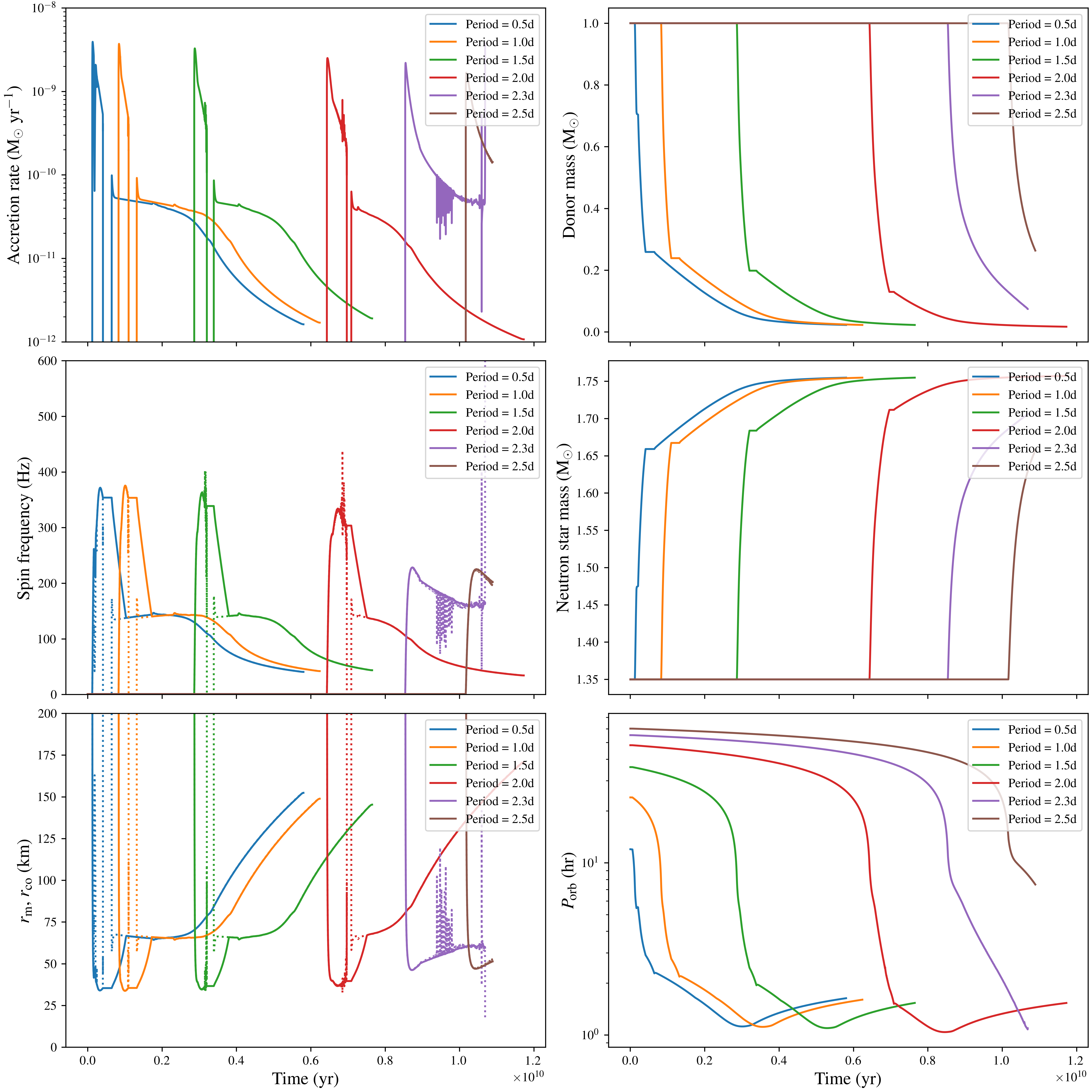}
  \end{tabular}}
   {\begin{minipage}[c][\textwidth][c]{\wd0}
    \usebox0
\caption{Evolution of parameters (accretion rate, companion star mass, spin frequencies [$\nu$, $\nu_{\rm eq}$], characteristic radii [$r_{\rm m}$, $r_{\rm co}$], neutron star mass, and binary orbital period $P_{\rm orb}$) for six selected values of initial $P_{\rm orb}$, as calculated using MESA. 
In each panel, other parameters are fixed at their canonical values (see Table~\ref{table1}). 
The dotted line represents $\nu_{\rm eq}$ and $r_{\rm m}$, while solid line is for $\nu$ and $r_{\rm co}$.  
These panels systematically show effects of initial $P_{\rm orb}$ on the evolution of various parameters
(section~\ref{subsec: detailed-effect}).} 

    \label{period days variation}
    \end{minipage}}
  \end{figure*}

\begin{figure*}
  \sbox0{\begin{tabular}{@{}cc@{}}
    \includegraphics[width=175mm]{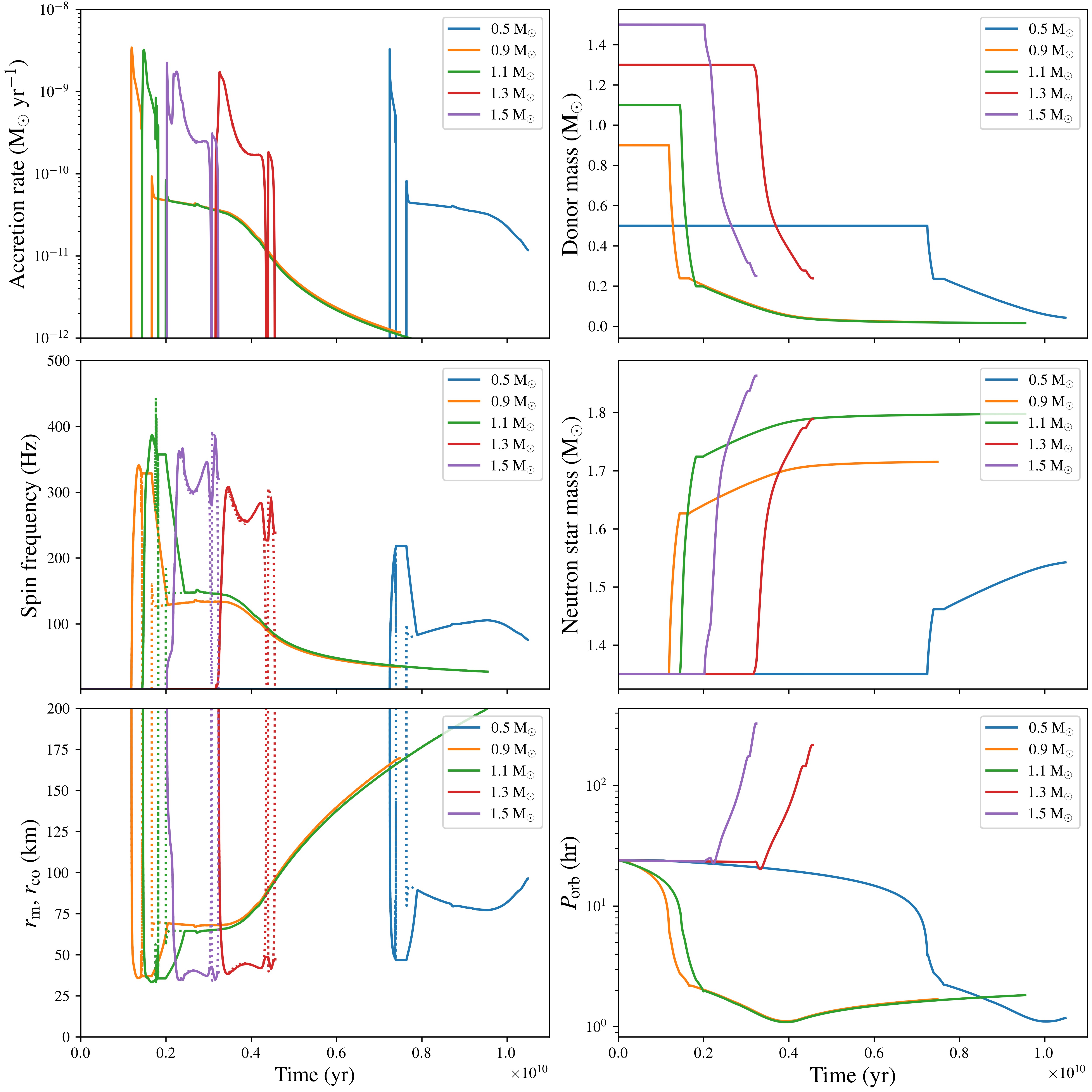}
  \end{tabular}}
   {\begin{minipage}[c][\textwidth][c]{\wd0}
    \usebox0
\caption{Similar to Figure~\ref{period days variation}, but the curves in each panel are for five selected initial companion star mass values, while initial $P_{\rm orb}$ and other parameters are fixed at their canonical values (see Table~\ref{table1}). 
These panels systematically show effects of initial companion star mass on the evolution of various parameters
(section~\ref{subsec: detailed-effect}).}
    \label{companion mass variation}
    \end{minipage}}
  \end{figure*}

\begin{figure*}
  \sbox0{\begin{tabular}{@{}cc@{}}
    \includegraphics[width=175mm]{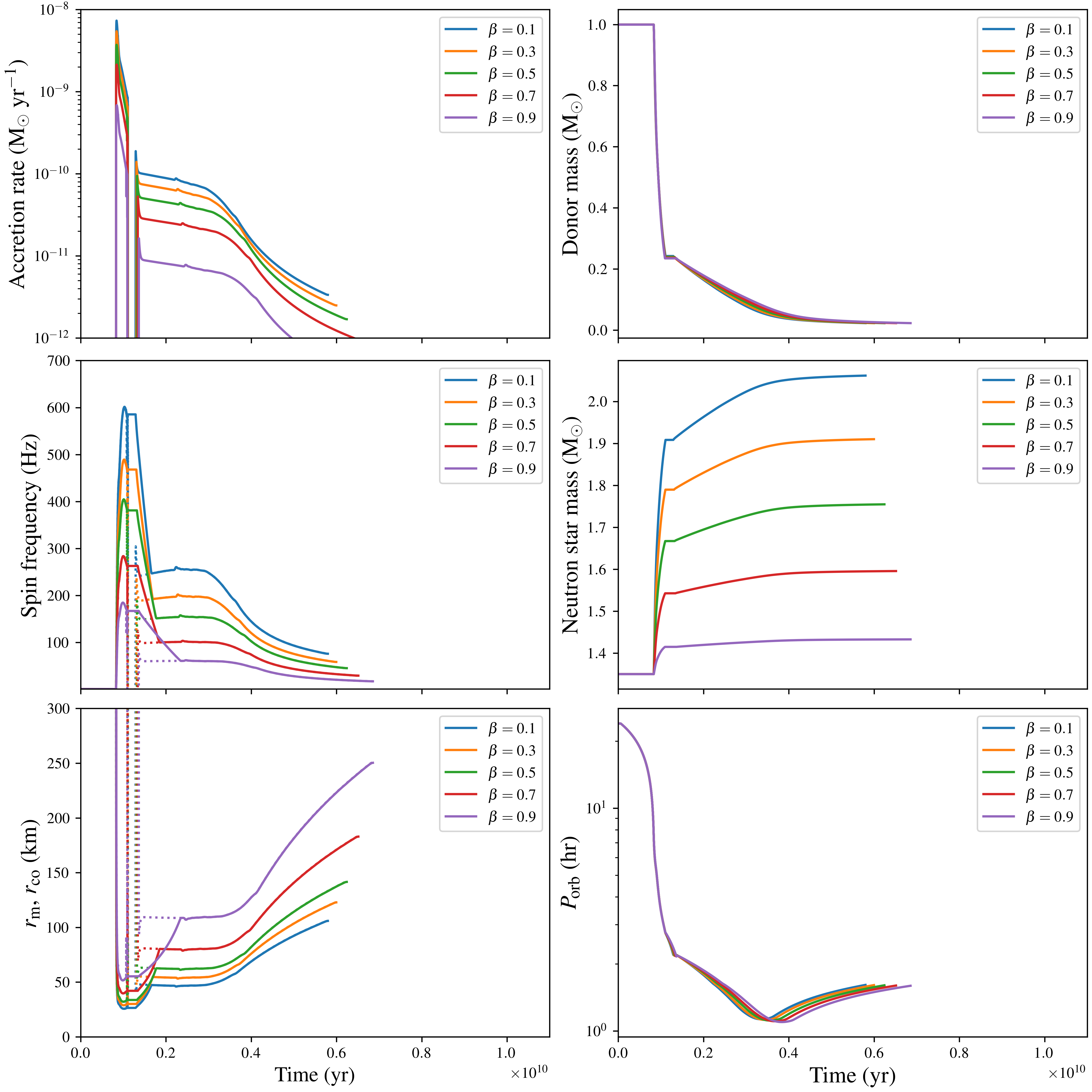}
  \end{tabular}}
{\begin{minipage}[c][\textwidth][c]{\wd0}
    \usebox0
\caption{Similar to Figure~\ref{period days variation}, but the curves in each panel are for five selected fractional mass loss $\beta$ values, while initial $P_{\rm orb}$ and other parameters are fixed at their canonical values (see Table~\ref{table1}). 
These panels systematically show effects of $\beta$ on the evolution of various parameters
(section~\ref{subsec: detailed-effect}).}
    \label{beta variation}
    \end{minipage}}
  \end{figure*}

In this paper, we systematically study the evolution trends of various NS LMXB along with exploring plausible evolutionary models for observed AMXPs using the MESA code.We particularly probe the long-term evolution of the NS spin frequency during accretion phase, which is poorly explored considering the realistic binary evolution, specifically using MESA. 
This can be useful to understand the distribution of known parameter values of MSPs, as well as the evolution of individual MSPs.
When the accreted matter falls on the NS, the new NS mass should be computed by converting the baryonic mass of this accreted matter into the gravitational mass.
We include this effect using the approximate formula  \ref{eq 12}, \citep{cipoletta} for Schwarzschild space-time (for simplicity), and show that the effect is not only substantial for $M_{\rm NS}$, but also significant for $\dot M$ and $\nu$ evolution (see Figure~\ref{cipoletta}). 
Particularly, if this effect is not considered, then the $M_{\rm NS}$ value can be overestimated by $\sim 10$\% resulting in an overestimate of $\nu$ evolution as well. 
In a few previous works \citep{jia&li_beta, goodwin}, it was argued that the mass transfer efficiency parameter $\beta$ does not affect the LMXB evolution significantly. Therefore, following \citet{podsiadlowski,chen} a canonical value of $\beta = 0.5$ was used. 
However, in the context of NS spin frequency evolution and some other parameters (such as NS final mass, characteristic radii $r_{\rm m}$, $r_{\rm co}$, etc.), $\beta$ plays an important role as it significantly affects the accretion rate onto the NS (see section~\ref{subsec:beta}). 
For instance, for very low values of $\beta$ (e.g., 0.1), the maximum $\nu$ value can be {high (see Figure~\ref{beta variation}) and possibly become more than 1 kHz for a relatively short duration at the initial part accretion, if the magnetic field values are adjusted accordingly.}. 
This implies the formation of a sub-millisecond pulsar, which has not been observed yet \citep{Patruno_2010NGW,submilisecond2,submilisecond1,cut_off2023}.
Another aspect is the total mass transfer to the NS, which depends not only on initial $M_{\rm comp}$(Figure~\ref{companion mass variation}) and initial $P_{\rm orb}$ (Figure~\ref{period days variation}), but also on $\beta$ (Figure~\ref{beta variation}).
For example, for $\beta = 0.5$, with all other parameters fixed at their canonical values, the total accreted mass to the NS is $\approx 0.4 M_{\odot}$. But, for $\beta = 0.1$,  the total accreted mass is found to be $\approx 0.71 M_{\odot}$, which is substantially higher than $0.46 M_{\rm \odot}$, as reported earlier \citep[eg.,][]{accreted_mass}. 
All of these parameters, play important role in obtaining the parameters of accreting sources i.e., AMXPs. 
We observe a correlation between initial $P_{\rm orb}$ and the age at which mass transfer starts, for a fixed initial $M_{\rm comp}$ value (Figure~\ref{period days variation}). 
But for a fixed initial $P_{\rm orb}$, we get this trend only for systems with initial $M_{\rm comp} \leq 1.0 M_{\rm \odot}$.
This observed pattern is consistent with the values of interaction time presented in table A1 of \citet{podsiadlowski}. 
The LMXBs can evolve into contracting or widening systems depending on the initial $P_{\rm orb}$ value being greater or less than the bifurcation period of the system, discussed in \citet{bifurcation, main_jia&li}. This bifurcation period can be substantially different for systems with different initial $M_{\rm comp}$.
For example, this period is $\sim 2$ days for initial $M_{\rm comp} = 1 M_{\rm \odot}$, whereas it is $\sim 0.8$ days for initial $M_{\rm comp} = 1.5 M_{\rm \odot}$, from our computations. This dependence of bifurcation period on companion mass helps us in setting up initial parameters for evolution models of AMXPs with very short $P_{\rm orb}$ but different types of companion stars.
If the companion is a WD, it implies that the system has been accreting for sufficient time, for the companion to evolve to the WD stage. Also, the initial companion mass should also be $\geq 1 M_{\rm \odot}$. In these cases, to get a convergent orbital evolution with small $P_{\rm orb}$ of a few hours, we set the initial $P_{\rm orb}$ according to the bifurcation period for the systems in that donor mass range.
Contracting systems show a higher rate of mass transfer and a longer duration of binary interaction than the widening ones, thus useful in obtaining the parameters of AMXPs. 




If the LMXB has a relatively large initial $P_{\rm orb}$ (e.g., $> 2$ days) for initial $M_{\rm comp} = 1.0 M_\odot$, most of the main sequence stellar evolution happens even before the mass transfer starts, thus limiting the duration of the accretion phase.
For these cases, the evolution ends at higher values of $\nu$ (around 300 Hz; see Figure~\ref{period days variation}). The companion star does not become fully convective for such cases usually, and hence the accretion does not halt due to a stop in magnetic braking, and the spin equilibrium is not temporarily broken during the LMXB phase. 





As described in section \ref{subsec: general example}, the evolution of frequencies ($\nu$, $\nu_{\rm eq}$) and  characteristic radii ($r_{\rm m}$, $r_{\rm co}$, $r_{\rm lc}$) are primarily governed by $\dot M$. 
We use the torque formulae mentioned in \citet{main_paper_sudip}, with different expressions of the disc-magnetospheric interaction torque for accretion and propeller phases.
Here is the general trend of the spin evolution.
Soon after the accretion starts, both $\nu$ and $\nu_{\rm eq}$ quickly rise to their maximum values, as accretion rate rises to a high value (e.g., $\gsim 10^{-9}$ M$_{\odot}$yr$^{-1}$).
The $\nu$ increases due to a positive torque (Equation~\ref{eq 19}) in the accretion phase.
But, after a small fraction of the LMXB duration, $\dot M$ decreases, and thus $\nu$ also decreases tracking $\nu_{\rm eq}$ 
(section~\ref{subsec: general example}). 
If the magnetic braking temporarily switches off in between, both $\dot M$ and torque become negligible for a small fraction of the LMXB duration.
During this time, $\nu$ decouples from $\nu_{\rm eq}$ and remains constant. 
After the accretion resumes, typically the NS is not in spin equilibrium, and spins down by a negative torque (Equation~\ref{eq 20}) in the propeller phase.
Then, the spin equilibrium is established again, and for most cases of our examples, as $\dot M$ declines, $\nu$ continues to track $\nu_{\rm eq}$ to a low value at the end of the LMXB phase.
Thus, at least for many realistic cases, the breaking from spin equilibrium at the last part of the LMXB phase \citep[e.g., ][]{Tauris-science} cannot explain the high $\nu$ values of radio MSPs.

However, as discussed in section \ref{sec:AMXPs}, we could reproduce $\nu$ values for some of the observed AMXPs. We demonstrate the reproduction of some main parameters for at least a few of the AMXPs. 
We find that for a few cases the companion is a BD or WD.
However, even though we provide possible evolutions for the parameter space of AMXPs with WDs, getting right NS spin for systems with BD companions (having $\nu$ as high as 600 Hz) along with other parameters like $P_{\rm orb}$ is challenging. This happens because our models have lower NS spin values in the later phase of LMXB evolution, i.e., when the companion reaches the BD state. 
But as shown by \citet{main_paper_sudip,sudip_2017_2,sudip2021}, if one considers the transient accretion (in the latter part of evolution due to $\dot M < \dot{M}_{\rm crit}$), $\nu$ is expected to approach a several times higher equilibrium spin frequency and thus could increase to a higher value (section~\ref{subsec: general example}). This could reproduce the observed spin frequency of AMXPs with BD companions. But one should also carefully consider initial parameters in such cases so that the spin-up is sustained at the end, along with the correct type of companion star and observed orbital period.
We plan to investigate such a complex spin evolution using MESA in a subsequent work.

\section{Summary}
\label{sec:summary}


In this paper, we investigate the long-term evolution of NS LMXB parameters, including the NS spin frequency, using the MESA code. Using our general LMXB evolution results, we also match the observed parameter space for accreting millisecond pulsars or AMXPs. This is useful to provide evolutionary models for AMXPs with their parameters,  including $\nu$, as well as to probe fundamental aspects of NSs. 
Here, we summarise our main findings and their implications. 
\begin{enumerate} 
\item  
We include the baryonic to gravitational mass conversion to calculate the NS mass evolution with a dynamic conversion ratio,
and find that its effects are crucial for realistic computation of some source parameters, including the NS spin frequency. 

\item 
We show that the effect of the fractional mass loss ($\beta$) from the system is crucial for the evolution and values of some parameters, including the NS spin frequency. 
We find that the total mass accreted to the NS critically depends on $\beta$ and could be higher than that previously suggested. We also show that NS magnetic field values have a crucial  effect on NS spin frequency and can be adjusted  within the expected range to match the observed spin frequency accordingly.

\item 
We find out the general nature of the complex evolution of the NS spin frequency, and how it is affected by the accretion rate, companion star magnetic braking, and other parameters.
A temporary switching off of magnetic braking may cause a halt of accretion and no change in the spin frequency for that period. 
We find that the spin frequency peaks at the early stage of the LMXB phase, and then typically declines, first relatively sharply, and then slowly, sometimes with alternate increase and decrease. This trajectory of spin evolution plays important role in determining possible models for observed AMXPs.

\item 
We predict possible long-term  evolutionary models (using MESA) for most of the observed AMXPs including their spin frequency evolution for the first time, to the best of our knowledge. 
However, it is challenging to match the high NS spin frequency for systems with WD/BD companions.
This problem could be solved if, considering findings of some previous papers \citep{main_paper_sudip,sudip2021}, the NS spin frequency increases towards the end of the LMXB phase due to transient accretion.


\item 
We find that the gravitational radiation typically does not have a significant effect on spin evolution during the persistent accretion phase.
But, as mentioned in some previous papers \citep[see ][for demonstration]{main_paper_sudip,sudip_2017_2},
this effect could be crucial for transient accretion.

\end{enumerate}



\section*{Acknowledgements}
Authors thank the MESA community forum for their help in addressing technical problems related to the  MESA code. Authors also thank the anonymous referee for constructive comments which helped improve the paper. SB acknowledges the support of the Department of Atomic Energy (DAE), India.

\section*{Data Availability}

The data of this paper have been generated using the publicly available “Modules for Experiments in Stellar Astrophysics” (MESA) code.
Any specific information and data related to this work will be made available on request.

\newpage
\newpage
\newpage


\newpage

\pagebreak
\bibliographystyle{mnras}
\bibliography{citations} 



\appendix



\bsp	
\label{lastpage}
\end{document}